\documentclass[final,1pt,times,twocolumn]{elsarticle}

\usepackage{amssymb}
\usepackage{braket}
\usepackage{amsmath}
\usepackage{amsfonts}
\usepackage{listings}
\usepackage{xcolor}
\usepackage{hyperref}

\newcounter{bla}

\journal{Computer Physics Communications}

\begin{document}

\begin{frontmatter}



\title{Efficient computation of optical excitations in two-dimensional materials with the Xatu code}


\author[a]{Alejandro Jos\'e Ur\'ia-\'Alvarez\corref{author}}
\author[a]{Juan Jos\'e Esteve-Paredes}
\author[a]{M. A. García-Blázquez}
\author[a,b]{Juan Jos\'e Palacios}

\cortext[author] {Corresponding author.\\\textit{E-mail address:} alejandro.uria@uam.es}
\address[a]{Departamento de F\'isica de la Materia Condensada, Universidad Aut\'onoma de Madrid, 28049 Madrid, Spain}
\address[b]{Instituto Nicol\'as Cabrera, Condensed Matter Physics Centre (IFIMAC), 28049 Madrid, Spain}

\begin{abstract}

Here we describe an efficient numerical implementation of the Bethe-Salpeter equation
to obtain the excitonic spectrum of semiconductors. This is done on the electronic structure calculated either at the simplest tight-binding level or through
density funcional theory calculations based on local orbitals. We use a simplified model
for the electron-electron interactions which considers atomic orbitals as point-like orbitals and a phenomenological screening. The optical conductivity can then be optionally computed within the Kubo formalism. Our results for paradigmatic two-dimensional materials such as hBN and MoS$_2$, when compared with those of more sophisticated first-principles methods, are excellent and envision a practical use of our implementation
beyond the computational limitations of such methods.

\end{abstract}


\end{frontmatter}



{\bf PROGRAM SUMMARY}

\begin{small}
\noindent
{\em Program Title:} Xatu                          \\
{\em CPC Library link to program files:} (to be added by Technical Editor) \\
{\em Developer's repository link:} \\ https://github.com/alejandrojuria/xatu \\
{\em Code Ocean capsule:} (to be added by Technical Editor)\\
{\em Licensing provisions:} GPLv3  \\
{\em Programming language:} C++, Fortran, Python              \\
{\em Nature of problem:} The exciton spectrum is obtained as the solution of the Bethe-Salpeter equation for insulators and semi-conductors. Constructing the equation involves determining the screening of the electrostatic interaction and then determining the matrix elements of the interaction kernel, which are computationally-intensive tasks, specially if one takes a purely ab-initio approach.\\
{\em Solution method:} The Bethe-Salpeter equation can be efficiently set up and solved assuming that the basis of the reference electronic structure calculation, obtained either from tight-binding models or density functional theory with actual localized orbitals, corresponds to point-like localized orbitals. This, in addition to using an effective screening instead of computing the dielectric constant, allows to obtain the interaction kernel at very low computational cost and, thereof, the exciton spectrum as well as the light absorption of materials.\\
{\em Additional comments including restrictions and unusual features:} 
The code requires using at least C++11, given that it uses version-specific features. All linear algebra routines have been delegated to the \texttt{Armadillo} library. \\

\end{small}

\section{Introduction}
\label{sec:intro}
Bound electron-hole pairs, namely excitons, are known to be largely responsible for the most prominent features of the optical response of semiconductors near the band edge\cite{Hanke1980}, particularly for low dimensional materials\cite{chernikov2014, cudazzo2011}. This includes, of course, absorption and photoluminescence, but also the photovoltaic response, where substantial efforts, both experimentally and theoretically, are being made for energy-harvesting real-life applications \cite{Mueller2018}. On the theory side, multiple ways to describe excitons have been developed varying in accuracy and sophistication \cite{quintela2022}, from an effective two-body description \cite{prada2015, berkelbach2013} and configuration interaction \cite{franceschetti1999, dvorak2019, bieniek2022} to many-body perturbation theory (MBPT) \cite{csanak1971, Strinati1988, louie2} or time-dependent techniques \cite{Onida2002, attaccalite2011, Ridolfi2020, Cistaro2023}. MBPT itself can be purely electronic or include electron-phonon interactions \cite{antonius2022, chen2020, chan2023}. The current standard for exciton calculations is GW-BSE: the GW approximation \cite{louie1} is used to correct the density functional theory (DFT) electronic band structure, specifically the gap in insulators and semiconductors \cite{Hedin_1999}. The resulting band structure is then used to compute the exciton spectrum with the Bethe-Salpeter equation (BSE). The accuracy of the MBPT approach has prompted the development of several software applications for the calculation of first-principles many-body excitations \cite{berkeleygw, yambo,exciting,Perfetto_2018}.

Being first-principles calculations, one can seek quantitative agreement with experiments, at the cost of computational time. Alternatively, one can seek less costly, qualitative comparison through an effective description of the interactions. Our code Xatu is intended for such purpose. While the base electronic structure can be computed at any degree of fidelity, from the simplest tight-binding (TB) model to the more sophisticated GW approximation, electron-hole interactions are taken into consideration through a simplified model where orbitals are considered to be point-like along with phenomenological models for screening. In principle, any band structure can be used as the starting point as long as it comes from a localized orbitals basis code; plane waves-based calculations are out of scope since they go against the nature of the approximation used for the interaction. Ultimately, these two approximations result in a considerable reduction of the computational cost.

Beyond the intrinsic speed-up coming from the calculation scheme itself, Xatu has been written mainly in \texttt{C++}, and is designed to be as efficient and general as possible while keeping its usability relatively simple. It targets a wide range of systems in the landscape of computational tools for optical excitations, from those that simply are out of the range of first-principles ones because of the complexity of the unit cell, to those that require a quick iteration, while obtaining qualitative and sometimes even quantitative agreement with experiments.

\section{Exciton theory}

\subsection{The Bethe-Salpeter equation}

Here we review the basic aspects of our theoretical approach, highlighting the simplifications and analogies with respect to the standarized GW-BSE method (see e.g. \cite{martin2016}).
From a quantum chemistry perspective, for the description of excitons we consider the exact, non-relativistic electronic Hamiltonian of the solid of interest:
\begin{equation}\label{Hamiltonian}
     H = H_0 + V =\sum_{i,j}t_{ij}c^{\dagger}_{i}c_{j} + \frac{1}{2}\sum_{i,j,k,l}V_{ijkl}c^{\dagger}_ic^{\dagger}_jc_lc_k,
\end{equation}
where the indices include orbital and position degrees of freedom; we restrict to basis of localized orbitals. $H_0$ describes the kinetic and ion-electron interaction terms and $V$ is the electrostatic interaction between electrons. Diagonalization of $H_0$ yields a Bloch eigenbasis $\ket{n\bold{k}}$ with energies $\varepsilon_{n\bold{k}}$, which here will correspond to insulating or semi-conducting materials. The interaction term in \ref{Hamiltonian} contains
\begin{align}\label{interaction_matrix_element}
    \nonumber V_{ijkl}&=\braket{i,j|V|k,l} \\ 
    &=\int d\bold{r} d\bold{r}' \varphi^*_i(\bold{r})\varphi^*_j(\bold{r}')V(\bold{r},\bold{r}')\varphi_k(\bold{r})\varphi_l(\bold{r}')
\end{align}
where $V(\bold{r},\bold{r}')$ is the two-body interaction. This can be the bare Coulomb interaction or some alternative interaction to take into account dimensionality or screening.
Since the non-interacting Hamiltonian $H_0$ describes insulating materials, it is usually a good approximation to take the ground state for the interacting Hamiltonian $H$ as the Fermi sea:
\begin{equation} \label{FermiSea}
    \ket{GS} = \prod_{n,\bold{k}}^{\varepsilon_{n\bold{k}}\leq \varepsilon_F}c^{\dagger}_{n\bold{k}}\ket{0}
\end{equation}
where $\ket{0}$ denotes the state with zero electrons, and $\varepsilon_F$ is the Fermi energy. Then, an electron-hole pair of center-of-mass momentum $\bold{Q}$ between the conduction band $c$ and the valence band $v$, and located at momentum $\bold{k}$ is defined as:
\begin{equation}
    \ket{v,c,\bold{k},\bold{Q}} = c^{\dagger}_{c\bold{k} + \bold{Q}}c_{v\bold{k}}\ket{GS}
\end{equation}
meaning that one electron of momentum $\bold{k}$ from the valence bands is promoted to the conduction bands with momentum $\bold{k}+\bold{Q}$. Note that even though we denote these states as electron-hole pairs, we are not actually using hole quasiparticle operators, but simply refer to the hole as the absence of an electron in the Fermi sea. We will stick to the electron picture throughout this work, unless specified otherwise. These electron-hole pairs will serve as the basis for the exciton states, $\ket{X}_{\bold{Q}}$:
\begin{align}
     \nonumber\ket{X}_{\bold{Q}} &= \sum_{v,c,\bold{k}}A_{vc}^{\bold{Q}}(\bold{k})\ket{v,c,\bold{k},\bold{Q}} \\
     &= \sum_{v,c,\bold{k}}A_{vc}^{\bold{Q}}(\bold{k})c^{\dagger}_{c\bold{k} + \bold{Q}}c_{v\bold{k}}\ket{GS}
\end{align}
Therefore, the exciton is expressed as a linear combination of electron-hole pairs over different bands and momenta. Note that $\bold{Q}$ serves as a good quantum number for the exciton states, since the interaction is momentum-conserving. The interaction only mixes electron-hole pairs with the same net momentum, which is $\bold{Q}$. This can be seen by computing explicitly a general interaction matrix element, $V_{ijkl}$.

Next, we determine the $A_{vc}^{\bold{Q}}(\bold{k})$ coefficients that minimize the expectation value $\braket{X|H|X}_{\bold{Q}}$:
\begin{equation}
    \frac{\delta E[X]}{\delta X} = \frac{\delta}{\delta X}\left[\frac{\braket{X|H|X}_{\bold{Q}}}{\braket{X|X}_{\bold{Q}}}\right] = 0
\end{equation}
Performing this derivative explicitly is equivalent to the problem of diagonalizing the Hamiltonian represented in the basis of electron-hole pairs:
\begin{equation}
\label{eigenproblem}
    \sum_{v',c',\bold{k}'}H_{vc,v'c'}(\bold{k}, \bold{k}', \bold{Q})A_{v'c'}^{\bold{Q}}(\bold{k}') = E_XA_{vc}^{\bold{Q}}(\bold{k})
\end{equation}
where $H_{vc,v'c'}(\bold{k}, \bold{k}', \bold{Q})=\braket{v,c,\bold{k},\bold{Q}|H|v',c',\bold{k}',\bold{Q}}$.
The expansion in electron-hole pairs of the exciton is actually an ansatz: we obtain exact eigenstates of the Hamiltonian restricted to a partition of the Hilbert space, $PHP$, where $P$ is a projector over the single electron-hole pairs.

\begin{equation}
    PHP = \sum_{\substack{v,c,\bold{k}\\ v',c',\bold{k}'}}H_{vc,v'c'}(\bold{k}, \bold{k}', \bold{Q})\ket{v,c,\bold{k},\bold{Q}}\bra{v',c',\bold{k}',\bold{Q}}
\end{equation}
In fact, if we only consider charge-conserving excitations, we could represent the Hamiltonian in the following way:
\begin{equation}
H = \bigoplus^{N_e}_{n=0} P_nHP_n + C, \ \text{with } 
\end{equation}
where 
\begin{align}
\nonumber &P_n=\sum_{
\substack{\{c_i\},\{v_i\}\\ \{c'_i\},\{v'_i\}}}\ket{\{c_i\}, \{v_i\}}\bra{\{c'_i\}, \{v'_i\}}\ 
\end{align}
and
\begin{align}
    \ket{\{c_i\}, \{v_i\}} = \prod^n_{i=1}c^{\dagger}_{c_i}\prod^n_{i=1}c_{v_i}\ket{GS}
\end{align}
$N_e$ is the total number of electrons, $C$ the coupling between the different excitation sectors, and $P_n$ is the projector over the n-th electron-hole pairs sector.
If instead of using the Bloch states from $H_0$ we formulate the problem in terms of the Hartree-Fock (HF) solution to \ref{Hamiltonian}, then the coupling between the Fermi sea and the single-pair sector, $P_0HP_1$, is exactly zero according to Brillouin's theorem\cite{szabo2012modern, gross1986many}. As we will mention later, we will assume that this always holds even when the ground state has not been calculated in the HF approximation. 
The same, however, is not true for $P_0HP_2$ or $P_1HP_2$, i.e., the interaction couples the ground state and the one electron-hole pair sector with the two electron-hole pairs sector. Thus, the proposed ground state and the exciton states are never exact but approximate eigenstates. Given that the material is insulating, we expect the coupling to be weak due to the energy differences, which justifies the ansatz. Keeping with the exact diagonalization approach, one could try to diagonalize the Hamiltonian including more excitation sectors. Although possible in principle, it becomes quickly unfeasible  since the Hilbert space in many-body systems grows exponentially and, in this case, the eigenstates would involve a mixture of excitations, losing the interpretation as a bound electron-hole pair.

Going back to \ref{eigenproblem}, we compute next the Hamiltonian matrix elements in the $H_0$ basis, which are given in terms of the single particle energies and the interaction matrix elements:
\begin{align}
    \label{h_matrixelement}
    \nonumber 
    &H_{vc,v'c'}(\bold{k},\bold{k}', \bold{Q}) = \\ \nonumber&\delta_{\bold{k}\bold{k}'}\delta_{vv'}\big[\varepsilon_{c\bold{k}+\bold{Q}}\delta_{cc'}+\Sigma_{cc'}(\bold{k}+\bold{Q},\bold{k}'+\bold{Q})\big]\\
 &-\delta_{\bold{k}\bold{k}'}\delta_{cc'}\big[\varepsilon_{v\bold{k}}\delta_{vv'}+\Sigma_{v'v}(\bold{k}',\bold{k})\big] -(D - X)_{vcv'c'}(\bold{k},\bold{k}',\bold{Q})
\end{align}
where
\begin{equation}
   \begin{split}
    D_{vc,v'c'}(\bold{k},\bold{k}', \bold{Q}) &= V_{c\bold{k}+\bold{Q},v'\bold{k}',c'\bold{k}'+\bold{Q},v\bold{k}} \\
     X_{vc,v'c'}(\bold{k},\bold{k}', \bold{Q}) & = V_{c\bold{k}+\bold{Q},v'\bold{k}',v\bold{k},c'\bold{k}'+\bold{Q}}
\end{split} 
\end{equation}
and
\begin{equation}
\begin{split}
\Sigma_{nm}(\bold{k},\bold{k}')=\sum_{j,\bold{k}''}^{\text{occ}}\big( V_{n \bold{k}, j\bold{k}'', m \bold{k}', j\bold{k}''} - V_{n\bold{k},j \bold{k}'', j\bold{k}'',m\bold{k}'}\big)
\end{split}
\end{equation}
$D$, $X$ correspond to the direct and exchange interactions between the electron-hole pair, whereas $\Sigma$ is the self-energy coming from the interaction of the electron/hole with the Fermi sea. At this point we could obtain the exciton spectrum diagonalizing (\ref{h_matrixelement}). Instead, it is more convenient to solve first for the ground-state of (\ref{Hamiltonian}) at the mean-field level, i.e. in the HF approximation \cite{isil2014}. If we now write (\ref{h_matrixelement}) in the HF band basis, we obtain:
\begin{align}
\label{bse}
    \nonumber (\varepsilon_{c\bold{k}+\bold{Q}} - \varepsilon_{v\bold{k}})A_{vc}^{\bold{Q}}(\bold{k}) + \sum_{v',c',\bold{k}'}K_{vc,v'c'}(\bold{k},\bold{k}',&\bold{Q})A_{v'c'}^{\bold{Q}}(\bold{k}') \\
    &= E_XA_{vc}^{\bold{Q}}(\bold{k})
\end{align}
where $\varepsilon_{n\bold{k}}$ are now the HF quasiparticle energies, and $K = - (D-X)$ is the interaction kernel. Thus, the self-energies are now incorporated into the quasiparticle energies instead. Note that the Fermi sea energy has been set to zero, so that exciton energies can be compared directly with the gap of the system. This is the standard form of the Bethe-Salpeter equation for excitons using the Tamm-Dancoff approximation (TDA) \cite{hirata1999, dreuw2005}, and it defines the starting point for any exciton calculation. The main difference with MBPT comes from the interaction kernel, which there involves a dynamically screened interaction, usually in the random-phase approximation \cite{hybertsen1985, adler1962, wiser1963}. The determination of the dielectric constant is a computationally intensive task \cite{berkeleygw}, which we avoid setting instead an effective static screening.

So far we have seen that it is more convenient to pose the exciton problem in terms of the HF basis, as it simplifies the problem and allows to decouple excitation sectors. 
In practice, we do not address the problem of determining the mean-field solution to (\ref{Hamiltonian}). Instead, we start directly from equation (\ref{bse}) assuming that the initial band structure, which is already known, verifies it. Namely, for tight-binding band structures we drop the self-energy terms assuming that we are using a HF solution. Alternatively, if the band structure comes from DFT or MBPT (e.g. GW approximation), then we also remove the self-energy terms since the quasiparticle energies already include self-energy corrections (although they do not cancel exactly with those from (\ref{h_matrixelement})). Thus, from now on we regard the starting band structure as the non-interacting Hamiltonian $H_0$.

\subsection{Interaction matrix elements}
With Eq. \eqref{bse} established, a practical expression for the interaction matrix elements (\ref{interaction_matrix_element}) remains to be obtained. The single-particle states, using a basis of localized orbitals, can be written as:
\begin{equation}
\label{single-particle-state}
    \varphi_{n\bold{k}}(\bold{r}) = \frac{1}{\sqrt{N}}\displaystyle\sum_{\bold{R}}e^{i\bold{k}\cdot\bold{R}}\sum_{i,\alpha}C^{n\bold{k}}_{i\alpha}\phi_{\alpha}(\bold{r}-\bold{R} - \bold{t}_i)
\end{equation}
where $\{\phi_{\alpha}\}$ denote the orbitals located at the atom $i$ of the motif and $N$ is the number of unit cells of the system. As mentioned before, this wavefunction may correspond to that of a tight-binding model (meaning that the spatial nature of the orbitals is ignored and are typically considered orthonormal), or a DFT calculation with a local orbital basis set, which are in general non-orthogonal. While the origin of the single-particle states can be different, for the actual calculation of the interactions we will treat them on the same footing, approximating them as point-like orthonormal orbitals.
Depending on how we treat the interaction, different working expressions for the matrix elements can be obtained. For instance, we address first the direct term, which is given by:
\begin{align}
    \nonumber &D_{vc,v'c'}(\bold{k},\bold{k}', \bold{Q}) \\
    &= \int d\bold{r}d\bold{r}'\varphi^*_{c\bold{k}+\bold{Q}}(\bold{r})\varphi^*_{v'\bold{k'}}(\bold{r}')V(\bold{r},\bold{r}')\varphi_{c'\bold{k'}+\bold{Q}}(\bold{r})\varphi_{v\bold{k}}(\bold{r}')
\label{integral}\end{align}
We substitute the single-particle Bloch states (\ref{single-particle-state}) in Eq. \eqref{integral}. Expanding each term, we end up having to evaluate the same four-body integral, but now between the orbitals that compose each state:
\begin{equation}
\label{fourbody}
    \int d\bold{r} d\bold{r}' \phi^*_{\alpha}(\bold{r})\phi^*_{\beta}(\bold{r}')V(\bold{r},\bold{r}')\phi_{\gamma}(\bold{r})\phi_{\delta}(\bold{r}')
\end{equation}
At this point, there are two ways to compute the present four-body integral: we can evaluate directly the interaction in real space, or, instead, use its Fourier series to work in reciprocal space. In both cases we consider point-like orbitals centered at $\bold{R}+\bold{t}_i$: 
\begin{equation}
\label{approximation}
    \phi_{\alpha}(\bold{r} - \bold{R} - \bold{t}_i)\phi_{\beta}(\bold{r} - \bold{R'} - \bold{t}_j) \approx \delta_{\alpha\beta} \delta(\bold{r} - \bold{R} - \bold{t}_i)\delta_{ij}\delta_{\textbf{R},\textbf{R}'}.
\end{equation}
Integrating in real space, after simplifying the resulting deltas, we obtain the following expression for the direct term $D$:
\begin{align}
\label{direct}
    \nonumber &D_{vc,v'c'}(\bold{k}, \bold{k}', \bold{Q}) \\
     &= \frac{1}{N}\sum_{ij}\sum_{\alpha\beta}(C_{i\alpha}^{c\bold{k}+\bold{Q}})^*(C_{j\beta}^{v'\bold{k}'})^*C_{i\alpha}^{c'\bold{k}'+\bold{Q}}C_{j\beta}^{v\bold{k}}V_{ij}(\bold{k}'-\bold{k})
\end{align}
where
\begin{equation}
    V_{ij}(\bold{k}'-\bold{k}) = \sum_{\bold{R}}e^{i(\bold{k}'-\bold{k})\bold{R}}V(\bold{R} - (\bold{t}_j - \bold{t}_i)).
\end{equation}
Here $V_{ij}(\bold{k}'-\bold{k})$ can be regarded as a lattice Fourier transform centered at $\bold{t}_j - \bold{t}_i$. Since it is defined as a sum over lattice vectors and not an integral, one cannot use the shift property from the Fourier transform. Attempting to do so would result in breaking the spatial symmetries of the Hamiltonian. Then, the direct term can be interpreted as the weighted average of the Fourier transform of the interaction between the electron and the hole, over all positions and orbitals. The exchange term is computed analogously:
\begin{align}
\label{exchange}
    \nonumber &X_{vc,v'c'}(\bold{k}, \bold{k}', \bold{Q})\\
    &= \frac{1}{N}\sum_{ij}\sum_{\alpha\beta}(C_{i\alpha}^{c\bold{k}+\bold{Q}})^*(C_{j\beta}^{v'\bold{k}'})^*C_{i\alpha}^{v\bold{k}}C_{j\beta}^{c'\bold{k}'+\bold{Q}}V_{ij}(\bold{Q})
\end{align}
In case that there is only one atom in the motif, then expressions (\ref{direct}), (\ref{exchange}) simplify even further since the interaction decouples from the tight-binding coefficients, yielding:
\begin{equation}
\begin{split}
    &D_{vc,v'c'}(\bold{k}, \bold{k}', \bold{Q}) = 
    \frac{1}{N}V(\bold{k}'-\bold{k})(U^{\dagger}_{\bold{k}+\bold{Q}}U_{\bold{k}'+\bold{Q}})_{cc'}(U^{\dagger}_{\bold{k}}U_{\bold{k}'})_{v'v}\\
    &X_{vc,v'c'}(\bold{k}, \bold{k}', \bold{Q}) = 
    \frac{1}{N}V(\bold{Q})(U^{\dagger}_{\bold{k}+\bold{Q}}U_{\bold{k}})_{cv}(U^{\dagger}_{\bold{k}'}U_{\bold{k}'+\bold{Q}})_{v'c'}
\end{split}
\end{equation}
where $U_{\bold{k}}$ is the unitary matrix that diagonalizes the Bloch Hamiltonian $H(\bold{k})$ \cite{FengchengWu2015}. The evaluation of these expressions is much faster than the corresponding ones \eqref{direct} and \eqref{exchange} for a general case. Additionally, for $\bold{Q}=0$, the exchange term (\ref{exchange}) becomes exactly zero, which is not true in general, although it is usually neglected. As mentioned before, for DFT band structures we evaluate the interaction using the same point-like approximation, performing first a Löwdin orthogonalization of the basis. This allows to improve the TB descriptions, incorporating fine details to the quasiparticle dispersion along the BZ. In such treatments, our interaction matrix elements are an approximation to the true ones involving ab-initio orbitals. Given that in DFT the orbitals are known (e.g. gaussian-type basis in the CRYSTAL \cite{crystal17} code), one could, in principle, evaluate the integrals (\ref{fourbody}) exactly for a closer ab-initio calculation of excitons.
  
The previous calculation corresponds to the evaluation of the interaction matrix elements
in real space. An alternative approach consists of writing the interaction as its Fourier series before evaluating (\ref{fourbody}) \cite{Cistaro2023, Ridolfi2020}:
\begin{equation}
    V(\bold{r}-\bold{r'})=\frac{1}{N}\sum_{\bold{q}}V(\bold{q})e^{i\bold{q}\cdot(\bold{r}-\bold{r'})}
\end{equation}
where
\begin{equation}
    V(\bold{q}) = \frac{1}{V_{cell}}\int_{\Omega}V(\bold{r})e^{-i\bold{q}\cdot\bold{r}}d\bold{r}
\end{equation}
$\Omega=NV_{cell}$ denotes the volume of the crystal. Usually, one takes $\Omega\to\infty$ meaning that we can evaluate the integral analytically, this is, $V(\bold{q})$ becomes the Fourier transform of the potential. Note, however, that $\bold{q}$ is not restricted to the first Brillouin Zone (BZ), and $V(\bold{q})$ is not periodic in the BZ. Therefore, in principle one has to sum over $\bold{q}\in\text{BZ}$, but also over reciprocal vectors $\bold{G}$, i.e.:
\begin{equation}
    V(\bold{r}-\bold{r'})=\frac{1}{N}\sum_{\bold{q}\in\text{BZ}}\sum_{\bold{G}}V(\bold{q} + \bold{G})e^{i(\bold{q} + \bold{G})\cdot(\bold{r}-\bold{r'})}
\end{equation}
The evaluation of the integral is done in the same way, although in this case there is a plane wave instead of the electrostatic interaction. This approach is particularly useful when using a plane wave basis, since it allows to evaluate the four-body integrals exactly without need for approximation (\ref{approximation}). The interaction matrix elements $D$, $X$ are now given by:
\begin{equation}
\begin{split}
\label{reciprocal_interaction}
    &D_{vc,v'c'}(\bold{k}, \bold{k}', \bold{Q}) = \frac{1}{N}\sum_{\bold{G}}V(\bold{k}-\bold{k}' + \bold{G})I^{\bold{G}}_{c\bold{k}+\bold{Q},c'\bold{k}'+\bold{Q}}(I^{\bold{G}}_{v\bold{k},v'\bold{k}'})^* \\
     &X_{vc,v'c'}(\bold{k}, \bold{k}', \bold{Q}) = \frac{1}{N}\sum_{\bold{G}}V(\bold{Q} + \bold{G})I^{\bold{G}}_{c\bold{k} + \bold{Q},v\bold{k}}(I^{\bold{G}}_{c'\bold{k'} + \bold{Q},v'\bold{k}'})^*
\end{split}
\end{equation}
where
\begin{equation}
    I^{\bold{G}}_{n\bold{k},m\bold{k}'} = \sum_{i\alpha}(C_{i\alpha}^{n\bold{k}})^*C_{i\alpha}^{m\bold{k'}}e^{i(\bold{k}-\bold{k}'+\bold{G})\cdot\bold{t}_i}
\end{equation}
Usually $V(\bold{q})$ decays fast enough so it suffices to sum only over $\bold{G}=\bold{0}$ for the excitons to converge in energy. Xatu allows to use the interactions evaluated in real-space (expressions (\ref{direct}), (\ref{exchange})), or in reciprocal space (expressions (\ref{reciprocal_interaction})). They are benchmarked in section \ref{sec:examples}; by default we use the interactions in real-space since the calculation converges faster with the number of points in the BZ mesh, it can be used rigorously for finite systems such as ribbons, and the current implementation performs on par with the reciprocal one. 

Once the interaction kernel is determined, Eq. \eqref{bse} can be solved to obtain the exciton energies and wavefunctions, i.e., the coefficients $A_{vc}^{\bold{Q}}(\bold{k})$. These can be used to compute different quantities. For instance, given that the exciton is written as a linear combination of electron-hole pairs with well-defined $\bold{k}$ quantum number, we can define the probability density of finding the exciton in a specific pair in $\bold{k}$-space as:
\begin{equation}\label{kwavefunction}
    |\psi_X(\bold{k})|^2=\sum_{v,c}|A_{vc}^{\bold{Q}}(\bold{k})|^2
\end{equation}
which is the straightforward definition since all electron-hole pairs are orthonormal to each other.

\subsection{Spinful excitons}

 If the single-particle basis includes spin, one can also compute the expected value of the total spin of the exciton, $S_z^T=S_z^e+S_z^h$. Given that we are using a fully electronic description of the exciton, we need to specify the electrons whose spin we want to measure. To this purpose, we write the total spin operator in second quantization as:
\begin{equation}
    S^T_z = \sum_{c',c,\bold{k}}\sigma_{c'\bold{k}+\bold{Q}, c\bold{k}+\bold{Q}}c^{\dagger}_{c'\bold{k}+\bold{Q}}c_{c\bold{k}+\bold{Q}} - \sum_{v',v,\bold{k}}\sigma_{v'\bold{k},v\bold{k}}c_{v'\bold{k}}c^{\dagger}_{v\bold{k}}
\end{equation}
where $\sigma_{nm}=\braket{n|S_z|m}$. The labels $c,c',v,v'$ refer exclusively to the conduction and valence bands used in the definition of the excitons. Note that the second term, which corresponds to the spin of the hole, has a minus sign. This is because holes, when described as quasiparticles, have opposite momentum and spin of the corresponding electronic state, i.e. $h^{\dagger}_{n,-\bold{k},-\sigma} = (-1)^{\sigma}c_{n\bold{k}\sigma}$, for states below the Fermi energy, $\varepsilon_{n\bold{k}} < \varepsilon_F$ \cite{fetter2012quantum}. These $h$ operators describe creation/annihilation of holes in terms of their electronic counterpart. Although we keep $\bold{k}$ the same (since we are still in the electronic picture), we already incorporate this minus sign to give a correct description of the total spin of the exciton. As we will see later, this sign change is also necessary to retrieve the known singlet and triplet states when summing angular momentum. The two pictures are equivalent, and all the previous calculations can be reproduced in the electron-hole picture.
The expected value of the total spin is then given by:
\begin{align}
\label{spin_total}
\nonumber \braket{X|S^T_z|X}=
\left[\sum_{v,c,\bold{k}}\sum_{c'}A^{\bold{Q}}_{vc}\right.&(\bold{k})(A^{\bold{Q}}_{vc'}(\bold{k}))^*\sigma_{c\bold{k}+\bold{Q}, c'\bold{k}+\bold{Q}} \\
     & \left. - \sum_{v'}A^{\bold{Q}}_{vc}(\bold{k})(A^{\bold{Q}}_{v'c}(\bold{k}))^*\sigma_{v\bold{k}, v'\bold{k}}\right]
\end{align}
If $[H_0, S_z] = 0$, then the spin projection $S_z$ is also a good quantum number for the Bloch states. Therefore, they can be written now as $\ket{n\bold{k}\sigma}$, or in real space as $\varphi_{n\bold{k}}(\bold{r})\chi_{\sigma}$, where $\chi_{\sigma}$ denotes the spin part of the state. This means that the spin operator $S_z$ is diagonal, $\sigma_{nm}=\sigma_{n}\delta_{nm}$, which allows us to simplify expression (\ref{spin_total}):
\begin{equation}
    \braket{S_z^T} = \sum_{v,c,\bold{k}}|A^{\bold{Q}}_{vc}(\bold{k})|^2(\sigma_{c} - \sigma_{v})
\end{equation}
Another consequence of having the spin well-defined is that it also propagates to the electron-hole pairs that serve as a basis for the exciton states, i.e. $\ket{\Tilde{v}, \Tilde{c}, \bold{k},\bold{Q}}=c^{\dagger}_{\Tilde{c}\bold{k}+\bold{Q}}c_{\Tilde{v}\bold{k}}\ket{GS}$, where $\Tilde{v}=(v,\sigma_v)$, $\Tilde{c}=(c,\sigma_c)$. In principle, we allow the spin of the electron and the hole to be different, $\sigma_c \neq \sigma_v$. Taking into account the spin in the computation of the interaction matrix elements, we arrive at constraints on which electron-hole pairs interact. Then the direct and exchange terms read:
\begin{equation}
\begin{split}
\label{spin_interaction_matrix_elements}
    D_{\Tilde{v}\Tilde{c},\Tilde{v}'\Tilde{c}'}(\bold{k},\bold{k}', \bold{Q}) &= \delta_{\sigma_c\sigma_{c'}}\delta_{\sigma_v\sigma_{v'}}D_{vc,v'c'}(\bold{k},\bold{k}', \bold{Q}) \\
    X_{\Tilde{v}\Tilde{c},\Tilde{v}'\Tilde{c}'}(\bold{k},\bold{k}', \bold{Q}) &= \delta_{\sigma_c\sigma_v}\delta_{\sigma_{v'}\sigma_{c'}}X_{vc,v'c'}(\bold{k},\bold{k}', \bold{Q})
\end{split}
\end{equation}
which can be directly obtained by substituting the single-particle states, since the spin part is not mixed with the orbital part of the states (i.e. $\ket{n\bold{k}\sigma}=\ket{n\bold{k}}\otimes\ket{\sigma}$). At this point, we can arrange the electron-hole pairs into four groups depending on their spin: 
$$\{\ket{++}, \ket{--}, \ket{+-}, \ket{-+}\}_e=\{\ket{\sigma_c\sigma_v}\}_e$$
The $e$ subindex is used to denote that this corresponds to the electronic picture. Then the Hamiltonian represented in terms of the spin groups, and taking into account (\ref{spin_interaction_matrix_elements}) becomes:
\begin{equation}
    H = 
    \begin{pmatrix}
        H_0 - D + X & X           & 0       & 0 \\
        X           & H_0 - D + X & 0       & 0 \\
        0           & 0           & H_0 - D & 0 \\
        0           & 0           & 0       & H_0 - D
    \end{pmatrix}
\end{equation}
where $H_0$, $D$, $X$ are blocks which include matrix elements corresponding to different electron-hole pairs but same spin group. If we now take into account that the hole in its quasiparticle representation must have spin opposite of that of the electron vacancy, then our states are $\{\ket{+-}, \ket{-+}, \ket{++}, \ket{--}\}_{eh}$, where $eh$ denotes electron-hole picture. Therefore, the exciton spectrum would be composed of groups of three triplet states and one singlet state, as when adding angular momenta. If instead we turn off the exchange interaction, then every state should have at least four-fold degeneracy. Any additional degeneracy would come from spatial symmetries of the Hamiltonian, in particular from the irreducible representations of the little group at $\bold{Q}$ (see \ref{appendix:symmetry}). 

\subsection{Real-space wavefunction}
Plotting the probability density \eqref{kwavefunction} is useful to extract some information about the exciton such as the wavefunction type ($s$, $p$, etc, following the hydrogenic model). The same can be argued for its real-space wavefunction, $\psi_X(\bold{r}_e, \bold{r}_h)$. However, obtaining it is not as straightforward as the $\bold{k}$ wavefunction. To do so, first we define the field operators as:
\begin{equation}
    \psi^{\dagger}(\bold{r}) = \sum_{n\bold{k}}\varphi^*_{n\bold{k}}(\bold{r})c^{\dagger}_{n\bold{k}},\ \psi(\bold{r}) = \sum_{n\bold{k}}\varphi_{n\bold{k}}(\bold{r})c_{n\bold{k}}
\end{equation}
where $\varphi_{n\bold{k}}(\bold{r})$ are the single-particle states in coordinate representation. Then, we can define the amplitude or real space wavefunction of the exciton in the following way:
\begin{equation}
    \psi_X(\bold{r}_e,\bold{r}_h) = \braket{GS|\psi(\bold{r}_e)\psi^{\dagger}(\bold{r}_h)|X}
\end{equation}
This definition is motivated by the fact that $\varphi_{n\bold{k}}(\bold{r})=\braket{GS|\psi(\bold{r})|n\bold{k}}$. Before computing the amplitude, it is convenient to switch to the electron-hole picture. The field operator written in terms of electron and hole operators is:
\begin{equation}
    \psi(\bold{r})=\sum_{c\bold{k}}\varphi_{c\bold{k}}(\bold{r})c_{c\bold{k}} + \sum_{v\bold{k}}\varphi_{v\bold{k}}(\bold{r})h^{\dagger}_{v\bold{-k}}\equiv \psi_e(\bold{r}) + \psi^{\dagger}_h(\bold{r})
\end{equation}
where $\psi_e(\bold{r})$, $\psi_h(\bold{r})$ are the annihilation field operator for the electrons, and holes respectively. Since we are switching from the electronic to the electron-hole picture, the same has to be done for the exciton state, $\ket{X}=\sum_{v,c,\bold{k}}A_{vc}^{\bold{Q}}(\bold{k})c^{\dagger}_{c\bold{k}+\bold{Q}}h^{\dagger}_{v,-\bold{k}}\ket{0}$. Evaluating the exciton amplitude in terms of the electron and hole field operators, we obtain:
\begin{align}
\label{real-space-wf}
    \nonumber \psi_X(\bold{r}_e,\bold{r}_h)&=\braket{GS|\psi_e(\bold{r}_e)\psi_h(\bold{r}_h)|X} \\ 
    &=\sum_{v,c,\bold{k}}A_{vc}^{\bold{Q}}(\bold{k})\varphi_{c\bold{k}+\bold{Q}}(\bold{r}_e)\varphi^*_{v\bold{k}}(\bold{r}_h)
\end{align}
To obtain the first equality note that there are four cross terms containing electron and hole field operators. Two of them are zero, since they they move around the electron or the hole [e.g. $\psi_e(\bold{r}_e)\psi^{\dagger}_e(\bold{r}_h)$], meaning that the final state is still orthonormal to the ground state. There is a third term consisting on creation of an electron and a hole, $\psi^{\dagger}_e(\bold{r}_e)\psi^{\dagger}_h(\bold{r}_h)$. This term is also zero because we assume that our ground state is the Fermi sea, meaning that it does not contain excited electrons. If this were the case, then the exciton could also consist on deexcitations or antiresonant transitions. This is known as the Tamm-Dancoff approximation, and it is also usually present in GW-BSE. To obtain the final expression for the wavefunction, it remains to substitute the expression of the field operators. One recovers the electron-hole pairs states of the exciton basis (up to a sign from operator permutation), and from orthonormality it results in expression (\ref{real-space-wf}).

At this point, to be able to plot the exciton real space wavefunction, we still need to evaluate (\ref{real-space-wf}) in terms of the single-particle states $\varphi_{n\bold{k}}(r)$. Since the exciton wavefunction depends on both the position of the electron and the hole, first we need to fix the position of either of them to be able to plot the wavefunction. Since we assume the orbitals are point-like, both the electron and hole can only be localized at the atomic positions, so we will evaluate the wavefunction and the probability density at these points only. 

We set the electron to be located at cell $\bold{R}_e$ and atom $\bold{t}_m$ of the motif, $\bold{r}_e=\bold{R}_e + \bold{t}_m$, while the hole is at position $\bold{r}_h=\bold{R}_h + \bold{t}_n$. Using the point-like approximation (\ref{approximation}), the probability density of finding the electron at a given position with the hole fixed reads:
\begin{equation}
    |\psi_X(\bold{R}_e + \bold{t}_m, \bold{R}_h + \bold{t}_n)|^2 = \sum_{\alpha\beta}|\psi^{\alpha\beta}_X(\bold{R}_e + \bold{t}_m, \bold{R}_h + \bold{t}_n)|^2
\end{equation}
where
\begin{align}
    \nonumber |\psi_X^{\alpha\beta}(\bold{R}_e + \bold{t}_m, &\bold{R}_h + \bold{t}_n)|^2 = \\
    \nonumber&\frac{1}{N^2}\sum_{v,c,\bold{k}}\sum_{v',c',\bold{k}'}A_{vc}^{\bold{Q}}(\bold{k})(A_{v'c'}^{\bold{Q}}(\bold{k}'))^*e^{i(\bold{k} - \bold{k}')\cdot(\bold{R_e} - \bold{R_h})}
    \\
    &\cdot C^{c,\bold{k}+\bold{Q}}_{m\alpha}(C^{c',\bold{k}'+\bold{Q}}_{m\alpha})^*(C^{v,\bold{k}}_{n\beta})^*C^{v',\bold{k}'}_{n\beta} 
\end{align}
For both the reciprocal and the real-space probability densities, one could expect them to have the symmetries of the crystal, since $[H,C]=0$, where $C$ is any symmetry operator from the space group. However, if the states are degenerate, then they are not necessarily eigenstates of the symmetry operators and consequently the associated densities will not be invariant under symmetry transformations. Still, in this case it is possible to define a probability density that preserves the symmetry of the crystal for each degenerate subspace:
\begin{equation}
\label{wf_invariant}
    |\psi_X(\bold{r}, \bold{r}_h)|^2=\sum_n|\psi_X^{(n)}(\bold{r}, \bold{r}_h)|^2
\end{equation}
where the index $n$ runs over exciton states degenerate in energy. An analogous expression holds for the $\bold{k}$ wavefunction. It is always good practice to check that the resulting probability densities preserve the symmetry of the crystal to ensure that the calculation was done correctly. The proof for the invariance under symmetry operations of (\ref{wf_invariant}) is given in \ref{appendix:symmetry}.

\subsection{Optical conductivity and light absorption}
As an example of a post-processing calculation, we investigate here the interaction of the material with an incident linearly-polarized electric field. We elaborate below how to compute the optical response by means of the exciton eigenfunctions.

For a sufficiently low-intensity and linearly-polarized homogeneus electric pulse, $\bold{\varepsilon}(t)$, the induced current per unit frequency in the bulk of the material can be written $J_a=\sum_{\beta}\sigma_{ab}(\omega) \tilde{\varepsilon}_{b}(\omega)$, where the linear optical conductivity reads \cite{pedersen2015}
\begin{equation}
\label{eq:excitonkubo}
\begin{split}
\sigma_{ab}(\omega)=&\frac{\pi e^2 \hbar}{V} \sum_{k}^{N_X} \frac{1} {E_{k}}\bigg[ V_{k}^{a}(V_{k}^{b})^\ast \bigg] \delta(\hbar\omega-E_{k})
\end{split}
\end{equation}
Here, $N_X$ is the number of exciton states, $E_k$ is the energy of the $k$-th excited state, $V_k^{a}=\braket{GS|\hat{v}^a|X_k}$ is the velocity matrix element (VME) transition to the ground state and $V$ is the volume of the solid under periodic boundary conditions. In the equation above, only excitons with $\bold{Q}=0$ are considered, as finite momentum excitons cannot be achieved by light incidence. Thus, we drop $\bold{Q}$ from the notation, and instead specify the excitation index $k$ in the exciton coefficients, $A^k_{vc}(\bold{k})$. The exciton VME is found to be 
\begin{equation} \label{eq:vme_ex}
    V_k^{a}=\sum_{cv \bold{k}}A_{vc}^{k}(\bold{k})v^{a}_{vc}(\bold{k}),
\end{equation}
where $v^{a}_{vc}(\bold{k}) \equiv \braket{v\bold{k}|\hat{v}^{a}|c\bold{k}}=i\hbar^{-1}\braket{v\bold{k}|[H_0,\hat{r}^a]|c\bold{k}}$ ($H_0$ is the non-interacting or mean-field Hamiltonian). With light polarized along the $a$ direction, an exciton is dark (or bright) if $V_k^{a}=0$ ($V_k^{a}\neq 0$). In general, the brightness of an exciton and its contribution to Eq. \eqref{eq:excitonkubo} is dictated by the selection rules of $A_{vc}^{k}(\bold{k})$ and $v^{a}_{vc}(\bold{k})$ over the Brillouin zone. The calculation of VMEs has to be worked out taking into account the underlying local basis of our approach. It is found  \cite{Esteve2023,lee2018} (we simplify the notation here by doing $i\alpha\rightarrow \alpha$ for the rest of the section)
\begin{equation} \label{eq:vme}
	\begin{split}	 &\braket{n\bold{k}|\hat{\bold{v}}|n'\bold{k}}=\\
   & \sum_{\alpha \alpha'} (C^{n\bold{k}}_{\alpha})^*C^{n'\bold{k}}_{\alpha'}\nabla_{\bold{k}} H_{\alpha \alpha'}(\bold{k}), \\
		&+ i\sum_{\alpha \alpha'}(C^{n\bold{k}}_{\alpha})^*C^{n'\bold{k}}_{\alpha'}\Big[ \epsilon_{n}(\bold{k}) \bold{\xi}_{\alpha \alpha'}(\bold{k})-  \epsilon_{n'}(\bold{k})\bold{\xi}_{\alpha' \alpha}^{\ast}(\bold{k}) \Big].
	\end{split}
\end{equation}
with $\bold{\xi}_{\alpha \alpha'}(\bold{k})=i\braket{u_{\alpha\bold{k}}|\nabla_{\bold{k}}u_{\alpha' \bold{k}}}$ the Berry connection between Bloch basis states.  After some algebra, it reads
\begin{equation} \label{eq:quantities}
\bold{\xi}_{\alpha \alpha'}(\bold{k})=\sum_{\bold{R}}e^{i\bold{k}\cdot \bold{R}}\braket{\alpha \bold{0}|\hat{\bold{r}}|\alpha' \bold{R}}+i\nabla_{\bold{k}}S_{\alpha \alpha'}(\bold{k}).   
\end{equation}
In the case of an underlying non-orthonormal local orbital basis, the overlap matrix $S_{\alpha \alpha'}(\bold{k})\equiv \braket{\alpha\bold{k}|\alpha'\bold{k}}$ is accounted and makes the Berry connection above non-hermitian. Instead, one has $\bold{\xi}_{\alpha \alpha'}(\bold{k})=\bold{\xi}_{\alpha' \alpha}^{\ast}(\bold{k})+i\nabla_{\bold{k}}S_{\alpha \alpha'}(\bold{k})$. Eq. \eqref{eq:vme} allows to evaluate the optical matrix elements by means of the non-interacting hamiltonian plus position matrix elements between the local orbitals.

In the case of an orthogonal basis set, as in tight-binding models, the overlap matrix is an unitary matrix at all points of the Brillouin zone. In this case VMEs read
\begin{equation} \label{eq:vmetba}
 v^{a}_{vc}(\bold{k})=\sum_{\alpha \alpha'}(C^{n\bold{k}}_{\alpha})^*C^{n'\bold{k}}_{\alpha'} \Bigg[\frac{\partial H_{\alpha \alpha'}(\bold{k})}{\partial k_a}+iH_{\alpha \alpha'}(\bold{k})(t^{a}_{\alpha'}-t^{a}_{\alpha}) \Bigg]
\end{equation}
This expression is sometimes known as the ``diagonal tight-binding approximation (TBA)'' in ab-initio calculations involving the maximally-localized Wannier functions \cite{pizzi2020,ibanez2018}, where the inter-orbital position matrix elements are discarded. Eq. \eqref{eq:excitonkubo} can thus be evaluated and is implemented in our code.

Additionaly, Eq. \eqref{eq:excitonkubo} can be compared with the frequency dependent expression for its non-interacting counterpart. In the limit of no correlations, it reduces to
\begin{align}
\label{eq:ipkubo}
   \nonumber \sigma_{ab}(\omega)=\frac{\pi e^2 \hbar}{V} \sum_{cv\bold{k}} \frac{1} {\varepsilon_{c \bold{k}}-\varepsilon_{v\bold{k}}}&\bigg[ v_{cv}^{a}(\bold{k})v_{vc}^{b}(\bold{k})  \bigg]\\
    &\cdot\delta(\hbar\omega-[\varepsilon_{c\bold{k}}-\varepsilon_{v \bold{k}}])
\end{align}
From the frequency-dependent optical conductivity one can obtain related quantities of interest. For instance, the ratio  of absorbed incident flux per unit frequency and unit length (considering vaccum surroundings) is \cite{wu2016} $S(\omega)=\sigma(\omega)/c\epsilon_0$, also called absorbance. Note that this quantity is ill-defined for 2D lattice systems. In such case, all the absorbance is assumed to occur at $z=0$ reference plane of the material.

\section{Implementation}
\label{section:implementation}

The programming languages of choice for the implementation of the exciton theory were \texttt{C++} and \texttt{Fortran}, which are the usual options for heavy numerical computations. In the case of \texttt{C++}, to facilitate manipulation of matrices we use the library \texttt{Armadillo} \cite{Sanderson2016}, on top of the usual libraries for linear algebra (\texttt{BLAS} and \texttt{LAPACK}). The core of the code was written in \texttt{C++}, except the post-diagonalization calculation of the optical conductivity being written in \texttt{Fortran}. This routine is wrapped inside the \texttt{C++} library. The software was designed with a hybrid approach in mind: previous packages such as DFT codes require the preparation of input files, which are then fed to the program and result in some output files which may be post-processed to extract information. We propose to use the same scheme, i.e. to prepare an input file which describes the system where we want to compute the excitons (namely the Hamiltonian $H_0$), and another one with the description of the excitons (participating bands, $\bold{k}$ mesh, etc). However, there is an alternative usage, which is employing directly the exciton API defined to built the program. This a common approach, where one builds a library to expose some functionality to the user (e.g. \texttt{Python} libraries). Therefore, one can define some system and script the computation of excitons using the API. This is advised whenever we are interested in performing some manipulation of the excitons, and not only obtaining the spectrum or the absorption. There is a third approach, consisting on using the system files to leverage the definition of the system to other programs (e.g. DFT), and then use the API instead of the exciton configuration file. The different forms to use the code will be reviewed in \ref{section:usage}. The CLI option parsing has been done using the header-only library \href{https://tclap.sourceforge.net/}{TCLAP}, which is distributed with this package.

\subsection{Complexity analysis}

Next we will discuss the numerical implementation of the exciton computation and related quantities. Solving the Bethe-Salpeter equation (\ref{bse}) amounts to diagonalizing the corresponding matrix $PHP$.
Diagonalization is done using the standard linear algebra libraries, meaning that the main problem is constructing $PHP$ as fast as possible. Consider a system formed by $N$ unit cells in total (meaning $\sqrt{N}$ along each Bravais vector for a two dimensional system). To treat the interaction rigorously, one has to compute the excitons on a BZ mesh with the same number of $\bold{k}$ points as unit cells, due to the periodic boundary conditions. Therefore, one has to compute $N^2$ matrix elements, and each of them requires computing the lattice Fourier transform, which involves summations over the $N$ unit cells. This has to be done for all possible band pairs $B$, so a naive implementation of (\ref{bse}) would have $\mathcal{O}(N^3B^2)$ time complexity, on par with matrix diagonalization algorithms. Note that each interaction matrix element also requires knowing the tight-binding coefficients $\{C^{n\bold{k}}_{i\alpha}\}$. If the dimension of the Bloch Hamiltonian is $M$, then diagonalizing on the fly the system for each element of the BSE would result in time $\mathcal{O}(N^2B^2(N + M^3))$.

The easiest way to reduce the time complexity of the BSE construction is to increase the space complexity, i.e. to precompute and store quantities that appear multiple times, instead of computing them on the fly. This can be done for the Bloch Hamiltonian eigenvectors. Before constructing $PHP$, we diagonalize $H(\bold{k})$ $\forall\bold{k}\in\text{BZ}$, and store the eigenvectors. At this point, if were to store all eigenvectors, the spatial complexity would go from $\mathcal{O}(1)$ to $\mathcal{O}(NM^2)$. Since we only need the eigenvectors corresponding to the bands that participate in the exciton formation, it suffices to store only those, meaning that the spatial complexity would be $\mathcal{O}(NMB)$, i.e. we have to store $N$ matrices of size $M\times B$. Accessing directly the eigenvectors results in a time complexity of $\mathcal{O}(N^3B^2 + NM^3)$. 

The same could be done for the lattice Fourier transform $V_{ij}(\bold{k}-\bold{k}')$. Since it depends on the difference between between two $\bold{k}$ points, we could simply store $V_{ij}$ for each pair of $\bold{k}$ points. This implies high spatial complexity $\mathcal{O}(N^2)$, but overall it does not report any speed advantage, since precomputing this would be of order $\mathcal{O}(N^3)$. However, it is possible to reduce the time cost of the algorithm: as long as the $\bold{k}$ point mesh covers the whole BZ uniformly (as given by Monkhorst-Pack), then we can map the $\bold{k}$ point difference back to a single $\bold{k}$ point using the periodicity of $V_{ij}(\bold{k}-\bold{k}')$:
\begin{align}
\label{periodic_k}
   \nonumber \forall \bold{k},\bold{k}'\in\text{BZ, } \exists \bold{G}\in\text{Reciprocal lattice, }& \bold{k}''\in\text{BZ}\\
   &\text{ s.t. } \bold{k}-\bold{k}' = \bold{G} + \bold{k}''
\end{align}
Therefore, it suffices to compute and store $V_{ij}(\bold{k})$ $\forall \bold{k}\in\text{BZ}$. Then, when initializing the matrix elements of $PHP$, one has to find the vector $\bold{k}''$ such that it verifies (\ref{periodic_k}). The time complexity now is $\mathcal{O}(N^2)$, which is a reduction of an order of magnitude. The space complexity is also reduced, being now $\mathcal{O}(N)$.

With this, the algorithm for determining $PHP$ has time order $\mathcal{O}(N^2B^2+NM^3)$, and the memory requirements are $\mathcal{O}(N + NMB)=\mathcal{O}(NMB)$. As we will see, this allows for very fast computation of the BSE matrix, meaning that the main bottleneck lies in the diagonalization, as it often happens. In some cases we might be interested in the whole spectrum, but usually it suffices to determine the lowest energy eigenstates. To address this, the code includes a custom implementation of the Davidson algorithm, which is suited to obtain the ground state of quantum chemistry Hamiltonians \cite{DAVIDSON1975}.
\\

So far the discussion was focused on how to reduce the complexity of the algorithm, but it is equally important to comment on how to perform the actual computation of the matrix elements. The big O notation neglects all constant factors, which is fine for theoretical considerations, but might have a considerable impact on the real behaviour of the code. The general strategy followed was to vectorize all calculations to make use of the highly optimized and parallel existing linear algebra routines. The remaining parts that do not allow vectorization, such as matrix element initialization in $PHP$, were all parallelized with \texttt{OpenMP}. Currently, all the parallelism is shared-memory and distributed parallelism might be implemented in the future.

For instance, consider the direct interaction term which requires computing expression (\ref{direct}). Supposed that the lattice Fourier transform of the interaction is already computed for all motif combinations $i, j$ and for all $\bold{k}$ points, we basically have to sum over tight-binding coefficients multiplied by the interaction. Given that the Bloch eigenstates are already stored as columns in matrices, we want to write this as matrix-vector products. Specifically, we can use $V_{ij}$ as a bilinear form, so with a well-defined matrix $\Tilde{V}$ the direct term can be written as:
\begin{equation}
    D_{vc,v'c'}(\bold{k},\bold{k}',\bold{Q}) = C^{T}_{cc'}\Tilde{V}(\bold{k}'-\bold{k})C_{v'v}
\end{equation}
where
\begin{equation}
    \Tilde{V} = V(\bold{k}-\bold{k}')\otimes \mathbb{I}_n, \ \ C_{nm} = C^*_n \odot C_m
\end{equation}
$\odot$ denotes element-wise array product, $C_n$ is the vector of coefficients corresponding to state $\ket{n}$ and $\mathbb{I}_n$ denotes a square matrix of ones of dimension $n$, $n$ being the number of orbitals per atom. Note that this expression is only valid if all atoms have the same number of orbitals. Otherwise, one must take into account the different number of orbitals per chemical species when performing the Kronecker´s products. The exchange term $X$ can be computed in an analogous way. Note that this assumes that the order of the single-particle basis is $\{\ket{i}\otimes\ket{\alpha}\otimes\ket{\sigma}\}$, i.e. for each atomic position, we run over orbitals, and for each orbital we run over spin. This is also relevant for the computation of the spin of the excitons, since it follows this convention.

As we mentioned at the beginning, to compensate for the lack of screening of the theory, one typically uses the Rytova-Keldysh potential \cite{rytova, keldysh} instead of the bare Coulomb potential in the context of two-dimensional materials. However, both interactions diverge at $r=0$. We regularize this divergence by setting $V(0)=V(a)$ \cite{FengchengWu2015}, where $a$ denotes the lattice parameter. Currently, the code only implements the Keldysh potential, given by:
\begin{equation}
\label{keldysh}
    V(r)=\frac{e^2}{8\varepsilon_0\Bar{\varepsilon}r_0}\left[H_0\left(\frac{r}{r_0}\right)- Y_0\left(\frac{r}{r_0}\right)\right]
\end{equation}
where $\Bar{\varepsilon}=(\varepsilon_m+\varepsilon_s)/2$, with $\varepsilon_s$, $\varepsilon_m$ being the dielectric constants of the substrate and the embedding medium (usually vacuum) respectively, and $r_0$ the effective screening length. Those three parameters have to be specified for all calculations. $H_0$, $Y_0$ are Struve and Bessel functions of second kind respectively.

Also, since the interaction decays quickly, we employ a radial cutoff, such that for distances $r>R_c$ we take the interaction to be zero. Then, the effective interaction is:
\begin{equation}
    \Tilde{V}(r)=\left\{\begin{array}{cc}
        V(a) & \text{if } r = 0 \\
        V(r) & \text{ if } r < R_c \\
        0 & \text{else}
    \end{array}\right.
\end{equation}
where $R_c$ is the cutoff radius. The cutoff has two purposes: first, it enforces the crystal symmetries in the transformed potential (as a function of $\bold{k}$). Secondly, it allows to compute the summation over lattice positions faster. Instead of evaluating the potential over all lattice positions, we restrict the sum to the lattice positions where we know the potential is different from zero. As for the interactions computed using the Fourier series of the potential, we set $V(\bold{q}=0)=0$ to remove the long wavelength divergence.
\\

Lastly, it is also worth mentioning how to compute the probability of finding the electron on a given spatial position (\ref{real-space-wf}). Since this requires two summations over $\bold{k}, \bold{k}'$, its cost would be $\mathcal{O}(N^2)$. To obtain the whole wavefunction, a priori we have to evaluate this over each position in the crystal, meaning that the cost would be $\mathcal{O}(N^3)$. However, this would be the worse case scenario in which the exciton is strongly delocalized in real-space. Usually, it will suffice to compute the real-space wavefunction on a contour of the hole position, for a few unit cells only. To actually compute the probability, we want to use the fact that we are storing the exciton coefficients as vectors. First, note that (\ref{real-space-wf}) can be written as:
\begin{align}
    \nonumber |\psi_X^{\alpha\beta}(\bold{t}_n + \bold{R}_e, &\bold{t}_m + \bold{R}_h)|^2 \\ 
    &= \left|\frac{1}{N}\sum_{v,c,\bold{k}}A_{vc}^{\bold{Q}}(\bold{k})e^{i\bold{k}\cdot(\bold{R_e} -\bold{R_h})}    
    C^{c,\bold{k}+\bold{Q}}_{m\alpha}(C^{v,\bold{k}}_{n\beta})^*\right|^2
\end{align}
which already reduces the complexity down to $\mathcal{O}(N)$. Then, the probability is computed as $||A\odot C||^2$, where $A$ is the vector of exciton coefficients that incorporates the exponential terms and $C$ are the tight-binding coefficients arranged such that they match the electron-hole pair ordering of the exciton.

\section{Examples}\label{sec:examples}

So far we have discussed the theory underlying the code and its numerical implementation. Therefore, it remains to show actual examples of the capabilities of the code. One context where excitons are relevant is valleytronics: materials with honeycomb structure which exhibit the band gap at the $\bold{K}, \bold{K}'$ points of the Brillouin zone (the "valleys"), and whose optical excitations can be tuned according to the valley \cite{mak2018, xiao2012}. The materials most commonly used for this purpose are transition metal dichalcogenides (TMDs), with formula $\text{WX}_2$, where W is the transition metal and S some chalcogenide. Another similar material that has become highly relevant is hexagonal boron nitride (hBN), although in this case due to its good properties as an insulating substrate \cite{Dean2010}.
These materials have become the prototypical examples to test the capabilities of an exciton code, and have been studied extensively. We will characterize the excitons in both hBN and $\text{MoS}_2$, i.e. obtain the exciton spectrum for $\bold{Q}=0$, show the associated wavefunctions and compute the optical conductivity. We will also show how a simple strain model of hBN can be used to break some crystal symmetries and modify the excitonic ground state.
All the calculations shown are done with the real-space approach to the interaction matrix elements and neglecting the exchange term, unless specified otherwise.

\subsection{hBN}

Monolayer hexagonal boron nitride has a large quasi-particle band gap, with ab-initio calculations predicting a value of $6-8$ eV depending on the method \cite{zhang2022}. As we will see, the band structure of hBN is relatively flat along the $\bold{M}-\bold{K}$ path in the Brillouin zone. This, in conjunction with small screening results in excitons that are strongly delocalized in reciprocal space, but are tightly bound in real space.

This material can be described easily with a minimal 2-band tight-binding model \cite{galvani}, equivalent to graphene but with opposite onsite energies for each atom of the motif. The tight-binding model for hBN reads:
\begin{equation}
    H= \sum_{i}\frac{\Delta}{2}(c^{\dagger}_ic_i-d^{\dagger}_id_i)+\sum_{\substack{<i,j> \\ i \neq j}}\left[ tc^{\dagger}_id_j + \text{h.c.}\right]
\end{equation}
where $c^{\dagger} (d^{\dagger})$ denote creation operators for B (N) atoms. The indices $i, j$ run over unit cells, and the summation over $<i,j>$ spans only the first-neighbours.
The parameters are $t=-2.3$ eV, $\Delta/2=3.625$ eV, and the corresponding system file can be found in the code repository under the folder \texttt{/models}.

\begin{figure}[h]
    \centering
    \includegraphics[width=1\columnwidth]{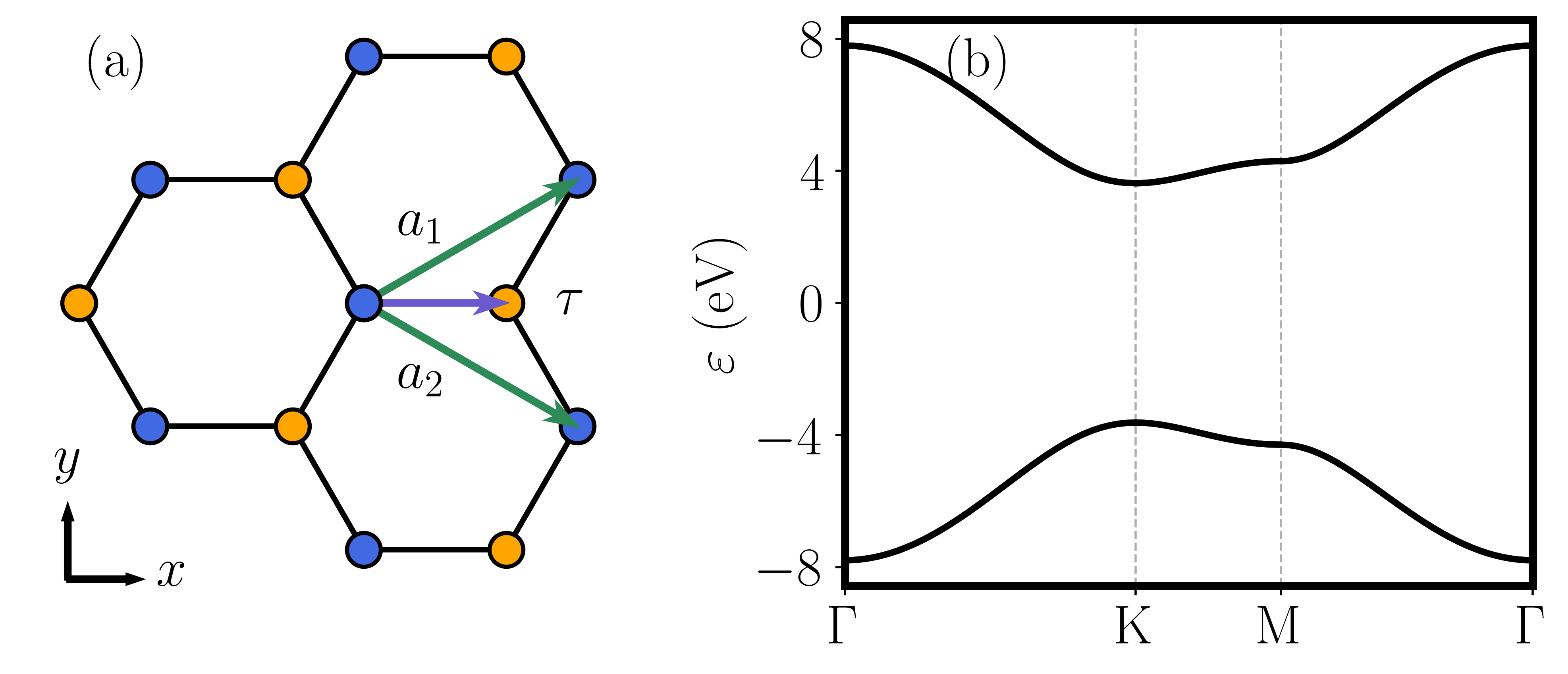}
    \caption{(a) hBN lattice and (b) band structure of the tight-binding model.}
    \label{fig:hbn_lattice}
\end{figure}

From a Slater-Koster perspective, hBN is described by $\text{p}_z$ orbitals. Taking the model to be spin-polarized for simplicity, there are only two bands and there must be only one electron per unit cell (half-filling) for it to be an insulator. Once the model is defined and the system file appropriately constructed, we can begin setting the parameters of the calculation. First, we need to specify the constants that appear in the Keldysh potential in Eq. (\ref{keldysh}). These parameters determine the strength of the electrostatic interaction and consequently affect the exciton binding energies. Here we follow previous works to set these quantities \cite{galvani}, but sometimes we will be interested in exploring the effect of tuning the dielectric constants, or instead we will want to set them to reproduce known experimental results. Nevertheless, values for typical substrates can be found in literature  and $r_0$ can also be estimated from ab-initio calculations \cite{prada2015, berkelbach2013}.

\begin{figure}[h]
    \centering
    \includegraphics[width=1\columnwidth]{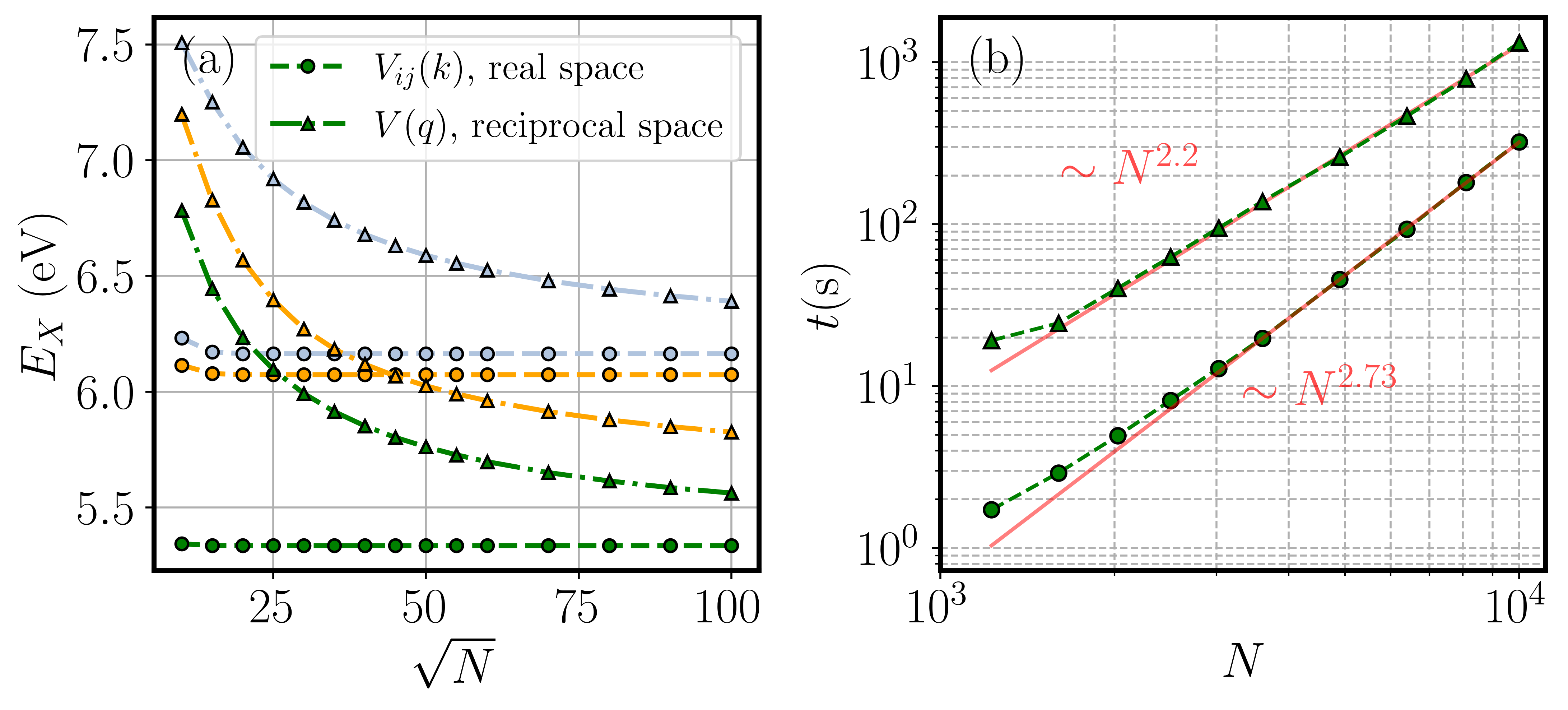}
    \caption{(a) Convergence of the ground state and first and second excited states as a function of the number of $\bold{k}$ points for hBN, computed both with interactions in real and reciprocal space. The reciprocal space calculations have to be converged also with respect to the number of reciprocal lattice vectors included, $N_G$, with $N_G = 25$ in this case. (b) Measured calculation time as a function of $N$. For both the real and reciprocal space calculations, the asymptotic behaviour is $\mathcal{O}(N^3)$. However, at small values of $N$ the required time is partially dominated by the BSE matrix initialization, which in both cases scales as $\mathcal{O}(N^2)$.}
    \label{fig:hbn_convergence}
\end{figure}

The other parameters of the exciton file are related to the convergence of the excitons themselves. Varying the number of $\bold{k}$ points in the mesh, $N_k$, one obtains the convergence curves shown in Fig. (\ref{fig:hbn_convergence}). The convergence has been done with both the default interactions (in real-space) and with reciprocal interactions (Fig. (\ref{fig:hbn_convergence}a)). For reciprocal interactions energies converge much slower than the real-space counterpart, on top of requiring summing over several $\bold{G}$ reciprocal cells. In materials with highly localized excitons in $\bold{k}$ space, it usually suffices to take only $\bold{G}=0$ (e.g.  $\text{MoS}_2$). However, we will see  later that hBN excitons are highly delocalized in reciprocal space, which is why the interaction can see neighbouring reciprocal unit cells.

After checking convergence, we can start studying the exciton themselves. The energies of the first 8 states and their degeneracies are given in table (\ref{tab:hbn_spectrum}). To make sense of the degeneracies, one has to check the character table of the point group of the material: hBN has the crystallographic point group $D_{3h}$, 
with both one- and two-dimensional irreducible representations.
Since the symmetry operations and their action on single-particle states are specific to each problem, the code does not address the problem of identifying the irreducible representation of each exciton, nor labeling them in terms of symmetry eigenvalues. Instead, we only check that the $\bold{Q}-$excitonic wavefunctions have the allowed degeneracies and \eqref{wf_invariant} is invariant under the little group at $\bold{Q}$.

\begin{table}[h]
    \centering
    \begin{tabular}{|c|c|c|c|}
        \hline 
        $n$ & Energy (eV) & Binding energy (eV) & Degeneracy \\
        \hline
        \hline
        1 & 5.3357 & -1.9143 & 2  \\
        2 & 6.0738 & -1.1762 & 1  \\
        3 & 6.1641 & -1.0859 & 2  \\
        4 & 6.1723 & -1.0777 & 1  \\
        5 & 6.3511 & -0.8989 & 2  \\
        \hline
    \end{tabular}
    \caption{Exciton spectrum from the tight-binding model for hBN computed with $N_k=60^2$. The binding energy $E_b$ is defined as $E_b=E_X - \Delta$, where $\Delta$ is the gap of the system.}
    \label{tab:hbn_spectrum}
\end{table}

\begin{figure}[t]
    \centering
    \includegraphics[width=1\columnwidth]{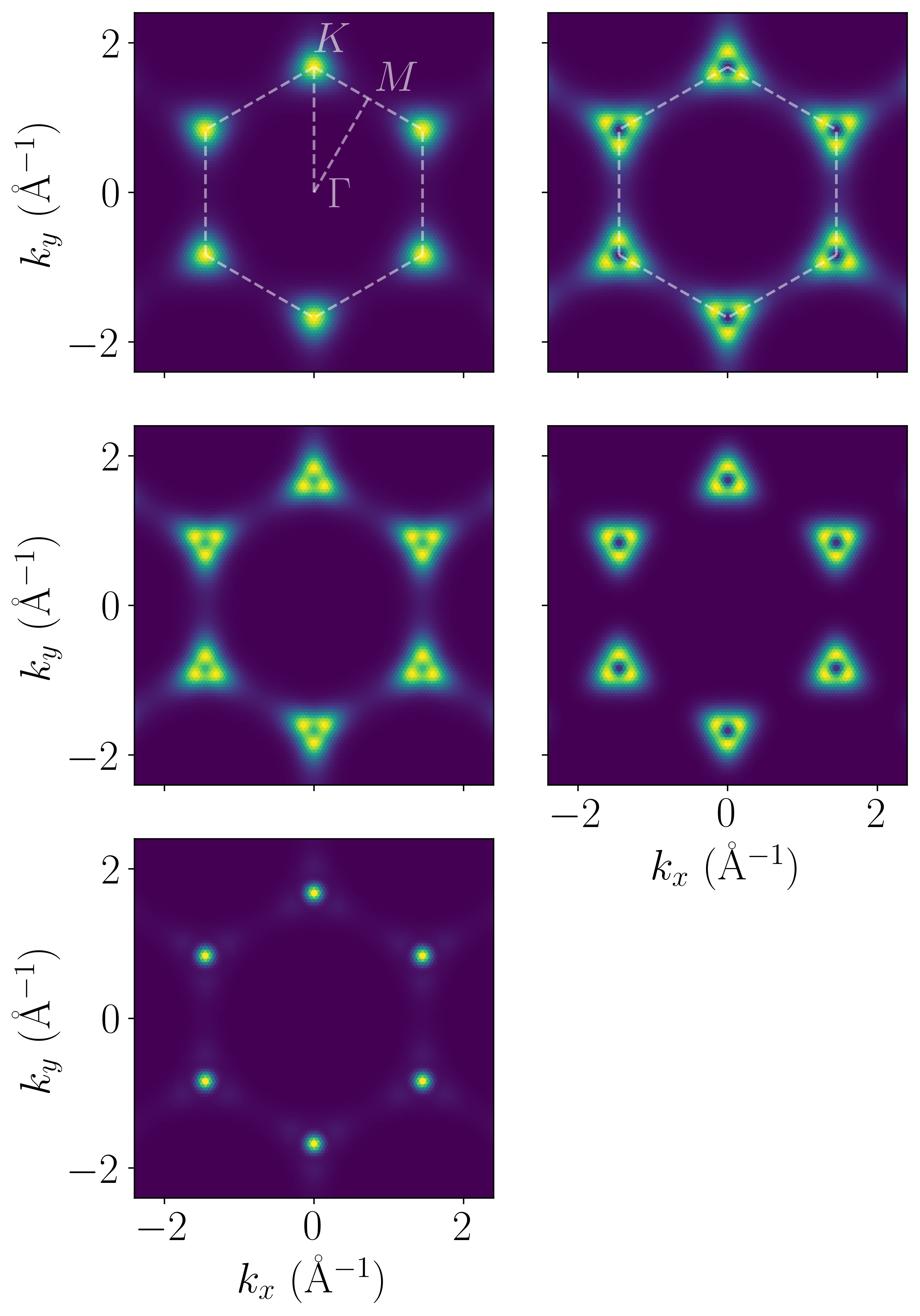}
    \caption{Plot of the $\bold{k}$ exciton probability densities in TB hBN, for the first 5 energy levels (from left to right, top to bottom). For each level, we actually show $|\Psi(\bold{k})|^2=\sum_n|\psi_n(\bold{k})|^2$ for $\bold{Q}=0$, where the index $n$ runs over degenerate states.}
    \label{fig:hbn_kwf}
\end{figure}

The $\bold{k}$ probability densities of the first eight excitonic states, grouped by degenerate levels, are shown in Fig. (\ref{fig:hbn_kwf}). Each energy level has the symmetry of the lattice, as expected since we are plotting (\ref{wf_invariant}). The additional symmetry in this case is due to time-reversal symmetry and the fact that $\bold{Q}=0$ is a time-reversal invariant momenta. We see that the wavefunctions peak at the valleys, although they also spread over the $\bold{K}-\bold{M}-\bold{K}'$ paths. This means that the excitons are formed by strongly interacting electron-hole pairs in $\bold{k}$ space, which explains why we need to sum over several reciprocal cells when using the reciprocal interactions. As for the shape of excitons, we find the common pattern: the first state peaks is $s$-like in the sense that it does not have nodes. The next state would be $p$-like and so on. Note that the hydrogen analogy only concerns the shape of the wavefunctions, and not the energy spectrum, which in general differs from the hydrogen series.\\

\begin{figure}[t]
    \centering
    \includegraphics[width=1\columnwidth]{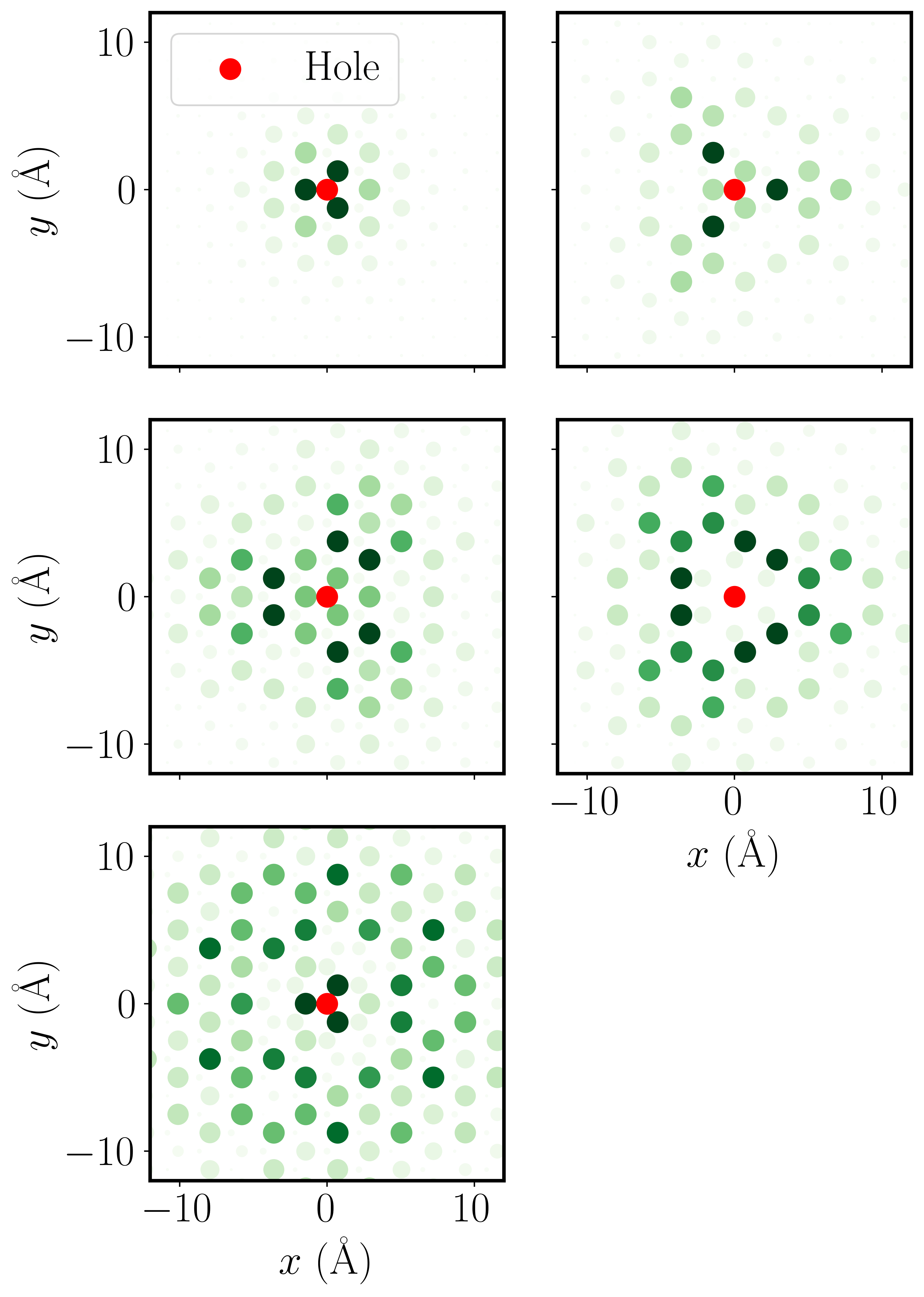}
    \caption{Plot of the real-space exciton probability densities in TB hBN with $\bold{Q}=0$, for the first 5 energy levels (from left to right, top to bottom). The red dot shows the position of the hole. For each level we plot the sum of the probability densities of each state of the degenerate subspace, $|\Psi(r_e, r_h)|^2=\sum_n|\psi_n(r_e, r_h)|^2$.}
    \label{fig:hbn_rswf}
\end{figure}

Since the excitons are delocalized in reciprocal space, we expect them to be strongly localized in real space. The real space densities of each degenerate level are shown in Fig. (\ref{fig:hbn_rswf}). The hydrogenic picture makes more sense when looking at the real-space wavefunction, since it can be understood then as the problem of two interacting opposite sign charges. The spectrum and the degeneracies do not match that of hydrogen, but the wavefunctions behave radially as we would expect. 

In hBN the spin-orbit coupling is small and it suffices to compute the excitons as a spinless system, in particular given that we are also neglecting the exchange interaction. If we consider a spinful system, without exchange again, we obtain exactly the same energy levels but now four-fold degenerate (on top of the previous spatial degeneracy). The same stands for both types of wavefunctions. 

\begin{figure}[t]
    \centering
    \includegraphics[width=0.9\columnwidth]{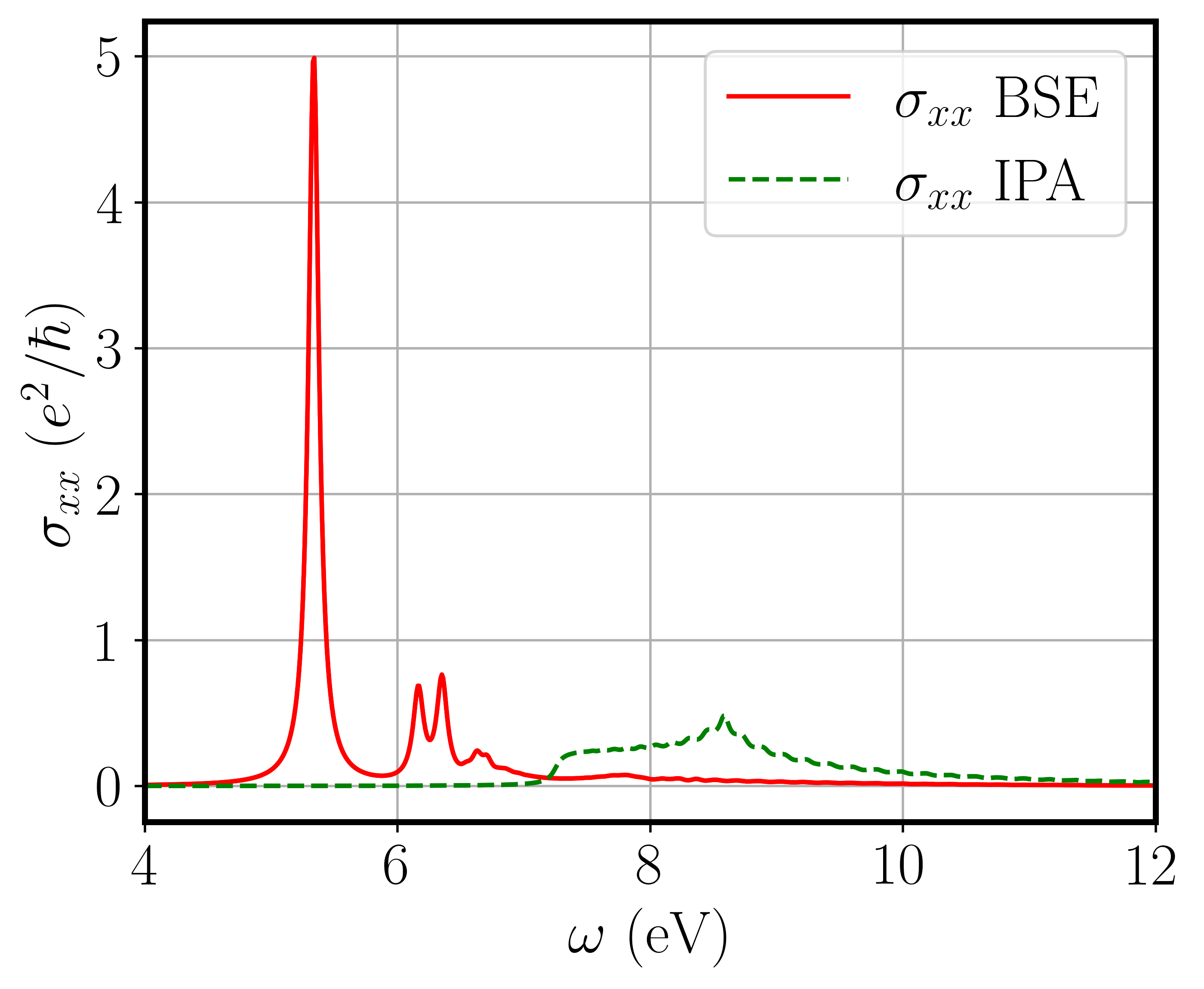}
    \caption{Optical conductivity of monolayer hBN as a function of the incident energy. We compare the conductivity obtained with the Kubo formula in the independent particle approximation (IPA), and with the BSE, which shows a dramatic change from the inclusion of excitons. }
    \label{fig:hbn_absorption}
\end{figure}

Our study of the exciton spectrum in hBN concludes with the calculation of the optical conductivity \cite{Ridolfi2020, pedersen2015, zhang2022}, which reflects the light absorbance from a source up to a constant factor. So far we have not discussed which excitons of the spectrum are bright or dark. This can be seen through the calculation of the optical oscillator strengths within Eq. \eqref{eq:excitonkubo}, which determine the transition rate for photon emission. The frequency-dependent conductivity of monolayer hBN is given in Fig. (\ref{fig:hbn_absorption}). Electron-hole interactions move the spectral power from the continuum to pronounced sub-band gap peaks. Attending to Table (\ref{tab:hbn_spectrum}) and Fig. (\ref{fig:hbn_kwf}), we see that non-degenerate excitons with mainly $s$ character are bright. The relative height of the peaks can be understood by looking at the magnitude of the wavefunctions near the $K$ and $K$' points. All bright excitons can be excited with linearly polarized light along two orthonormal polarization directions, giving rise to an isotropic conductivity consistent with the $D_{3h}$ point group of the material.

It is of interest to check the validity of the results against a more refined description of the band structure of the material. This can be done with the code by using a local orbital-based DFT calculation as the starting Hamiltonian, instead of using a parametrized tight-binding model. The exciton energies will depend on the gap as estimated from the functional used, but we expect to get similar wavefunctions and conductivity. Since we consider several orbitals for each chemical species now, we have multiple valence and conduction bands so we should converge the exciton with respect to the number of bands as well. It is a proper check to do, but in this case the different bands are well separated, so their effect should be negligible.

\begin{figure*}[h]
    \centering
    \includegraphics[width=0.8\textwidth]{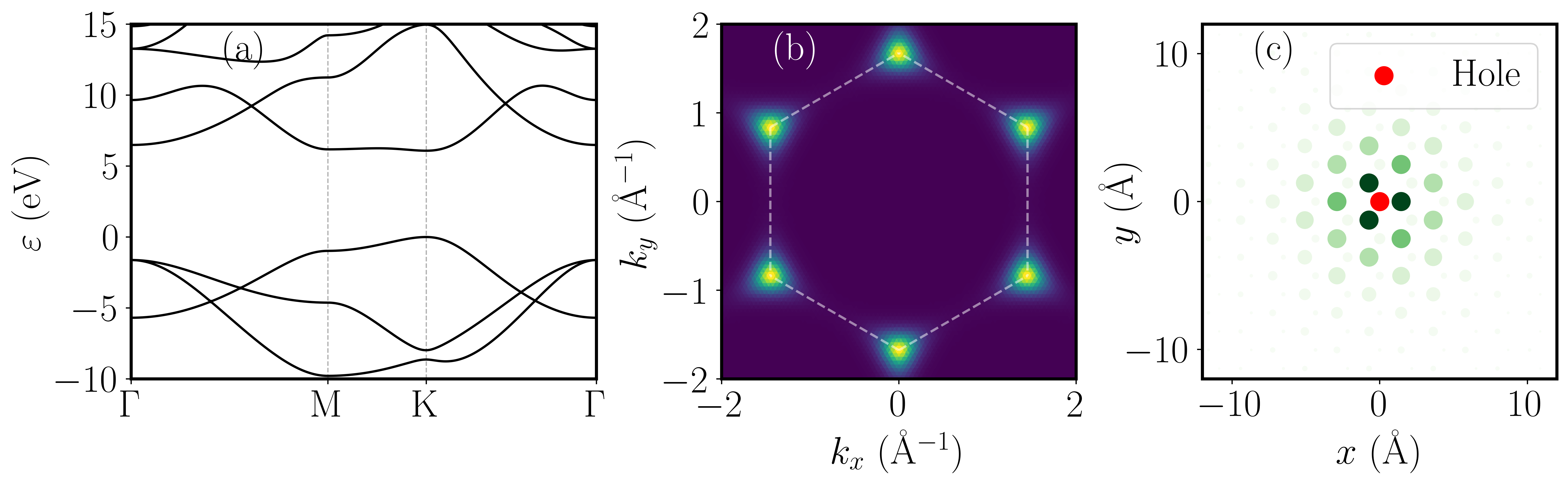}
    \caption{(a) DFT band structure of monolayer hBN as obtained with the HSE06 functional. (b) Reciprocal and (c) real-space probability densities of the $\bold{Q}=0$ ground state exciton with $N_k=60^2$, $N_c=N_v=1$. The DFT calculation involved a basis size of 36. We have run successfully exciton calculations on different systems with varying basis sizes, from 8 to 92.}
    \label{fig:hbn_dft}
\end{figure*}

The DFT band structure and the wavefunctions of the ground state exciton are shown in Fig. (\ref{fig:hbn_dft}). One could use standard LDA functionals, but here we opt for a hybrid functional (HSE06\cite{krukau2006} in this case), which is efficiently implemented in CRYSTAL \cite{crystal17}. This type of functional yields a better estimation of the single-particle gap due to a different treatment of the exchange-correlation term. For both LDA (not shown) and hybrid functionals such as HSE06, the wavefunctions closely resemble those obtained with TB models. For instance, we observe the same sublattice polarization present in the TB real-space densities with the HSE06 calculation (Fig. \ref{fig:hbn_dft}c).
The energy spectrum shows the same degeneracies, although the positions of some of the levels are exchanged. 

 To illustrate the applicability of the code beyond standard cases, we now study the effect of strain on the exciton spectrum. If we apply some uniaxial in-plane strain along the $x$ axis, the point group of the material will change to $C_{2v}$ (with rotation axis along $x$). The degeneracy of the ground state came from the spatial symmetries, meaning that it should be broken for any strain value, given that all irreducible representations of $C_{2v}$ are of dimension 1. Therefore, we can study the energy splitting of the ground state as a function of the applied strain.

\begin{figure}[h]
    \centering
    \includegraphics[width=0.5\columnwidth]{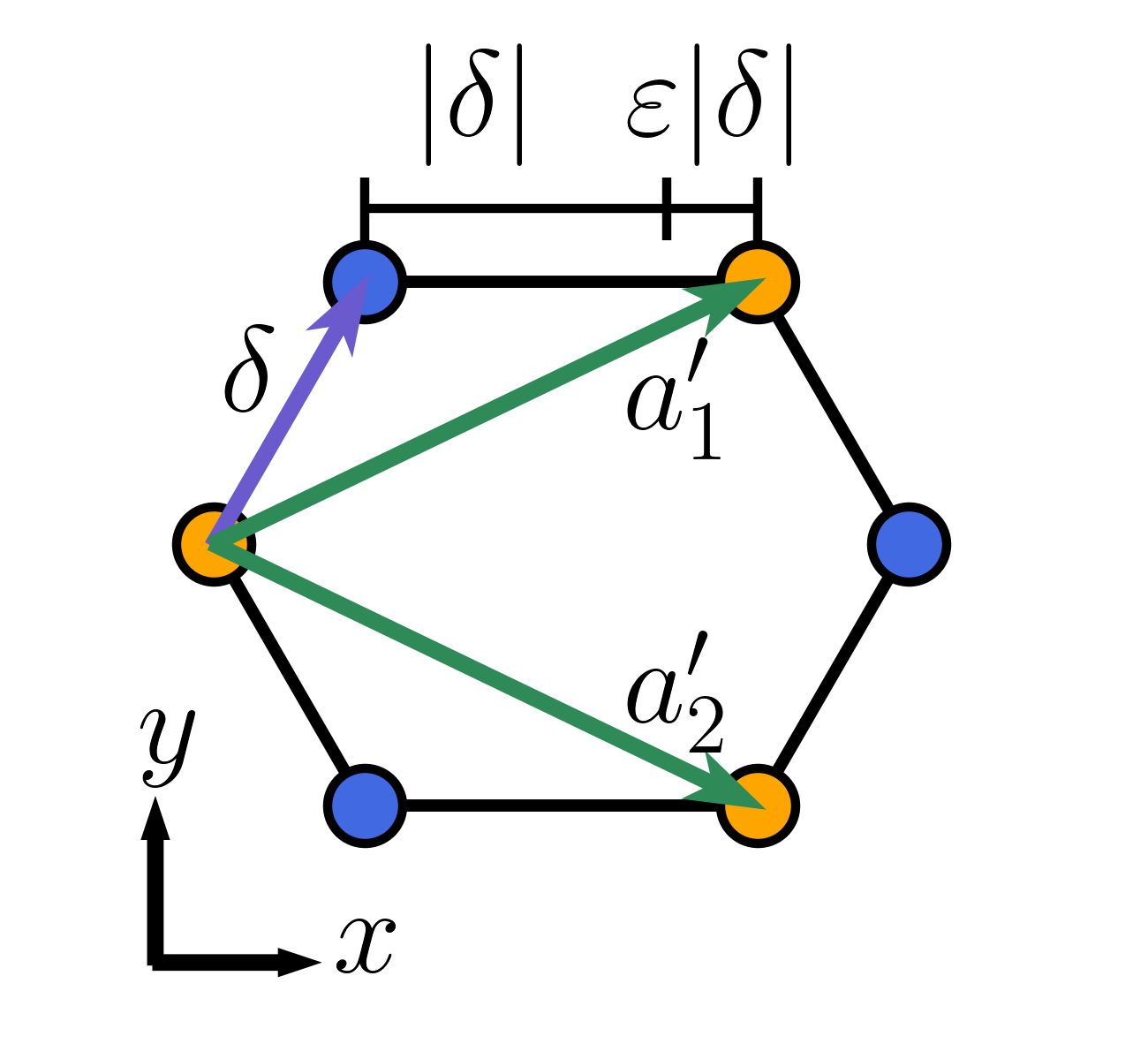}
    \caption{Schematic of the distortion of the hBN lattice due to the application of uniaxial strain along the $x$ axis. Note that in reality all bonds should be distorted, due to the phenomena of Poisson contraction.}
    \label{fig:hbn_strain}
\end{figure}

The strain model used is fairly straightforward. Based on the original tight-binding model, we now consider the hopping parameters to have an exponential dependence on the distance:
\begin{equation}
    t(r) = t_0e^{-a(r-r_0)},
\end{equation}
where $a$ is some decay length, $t_0$ the original value of the hopping and $r_0$ the reference length. Additionally, the distortion of the lattice due to strain is taken to affect only bonds parallel to the strain. A rigorous approach would have to implement appropriate distortion of all atomic positions according to the stress tensor \cite{nuno_strain}, but for our purposes this simple model suffices. This is illustrated in Fig. (\ref{fig:hbn_strain})

\begin{figure}[h]
    \centering
    \includegraphics[width=1\columnwidth]{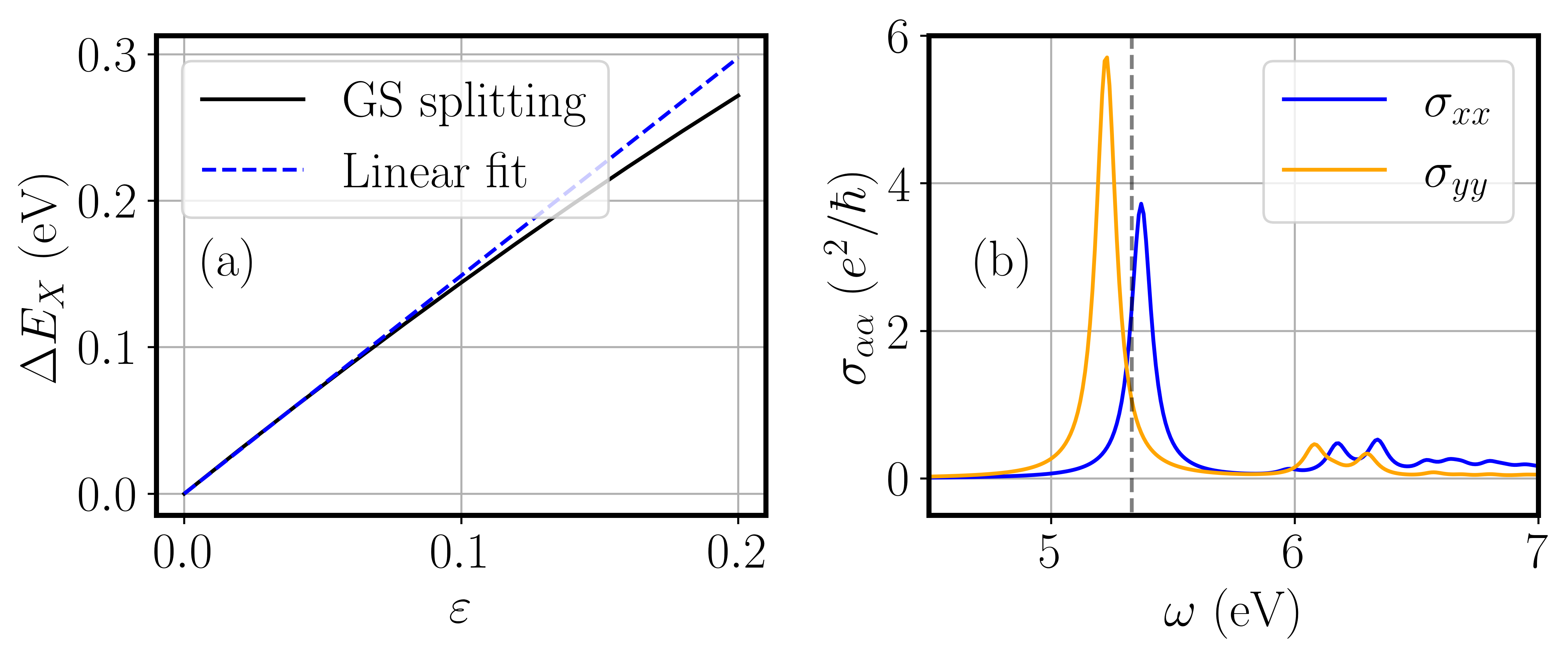}
    \caption{(a) Energy splitting of the ground state exciton as a function of strain in hBN. We observe that the splitting is linear on strain, for small values. (b) Frequency-dependend conductivity on strained hBN, $\varepsilon=0.1$. The dashed line shows the position of the ground state for $\varepsilon=0$.}
    \label{fig:hbn_strain_exciton}
\end{figure}

The procedure to study the exciton spectrum as a function of strain is as follows: we generate different system files (i.e. different Hamiltonians) for different values of the strain, which translates into different atomic positions. Then, we run the exciton simulation for each system file, storing the energies. As we expected, now all states are non-degenerate because of the symmetry group $C_{2v}$. We can plot the ground state splitting as a function of strain, which is shown in Fig. \ref{fig:hbn_strain_exciton}(a). In Fig.\ref{fig:hbn_strain_exciton}(b) we show the conductivity for some finite value of the strain. The response is no longer isotropic due to the lattice symmetry breaking caused by strain, where the exciton peaks shift for both light polarizations.

\subsection{$\text{MoS}_2$}

\begin{figure}[h]
    \centering
    \includegraphics[width=1\columnwidth]{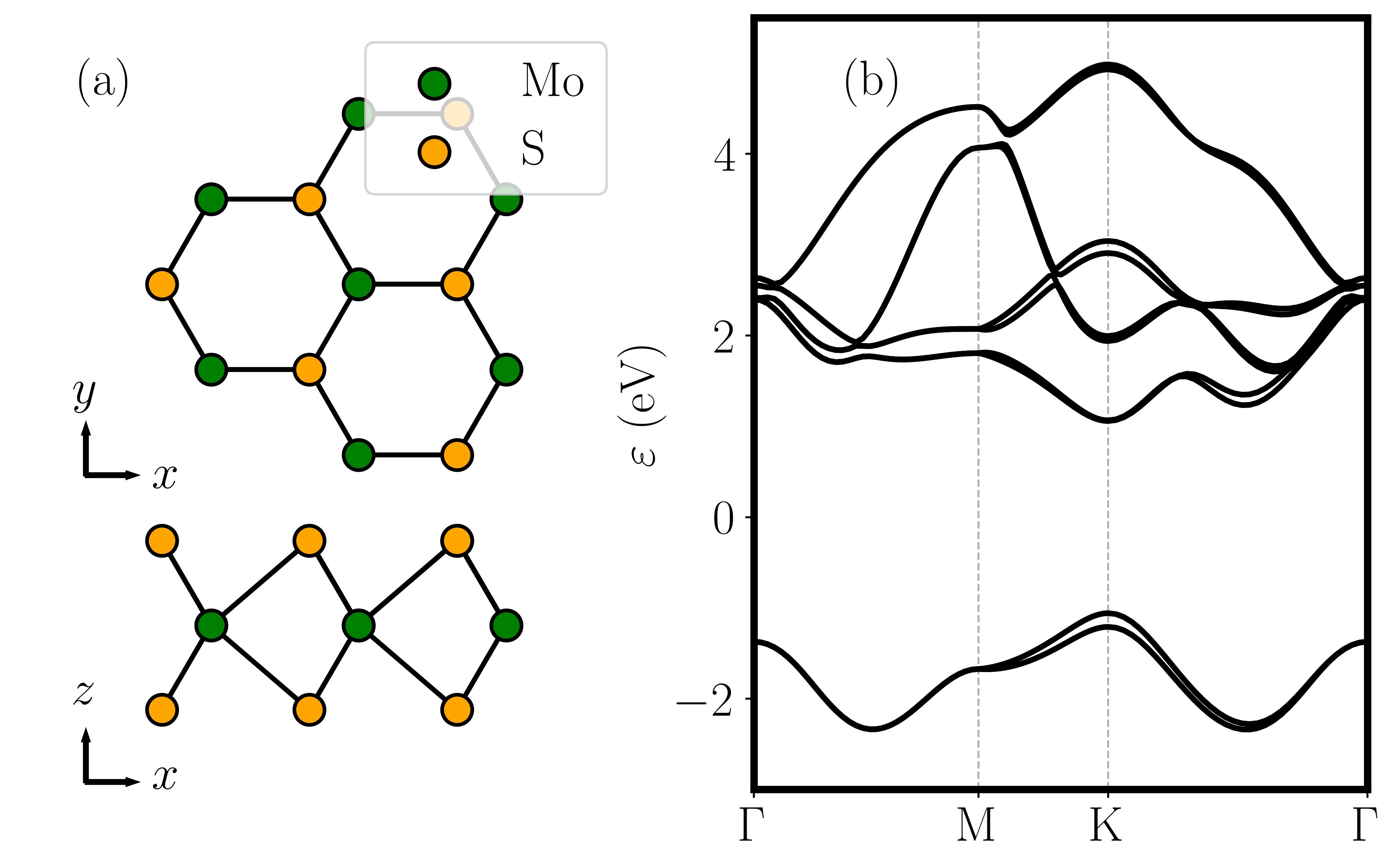}
    \caption{(a) Crystal and (b) tight-binding band structure of $\text{MoS}_2$.}
    \label{fig:mos2_bands}
\end{figure}

To conclude the examples section, we also analyze the exciton spectrum of $\text{MoS}_2$. Same as hBN, in monolayer form this material presents itself in a honeycomb lattice, although it is not planar. Instead, it is formed by three layers of composition S-Mo-S respectively. 
The description of the band structure of $\text{MoS}_2$ requires a more complex model, which is why we use it to showcase the code. We use a Slater-Koster tight-binding model \cite{Ridolfi_2015}, where each chemical species has a different set of orbitals (Mo has $d$ orbitals, and S only $p$ orbitals). This, together with the non-negligible spin-orbit coupling results in a more complex band structure than that of hBN. Both the lattice and the band structure can be found in Fig. (\ref{fig:mos2_bands}).

\begin{table}[h]
\label{tab:mos2}
    \centering
    \begin{tabular}{|c|c|c|c|}
        \hline 
        $n$ & Energy (eV) & Binding energy (eV) & Degeneracy \\
        \hline
        \hline
        1 & 1.7673 & -0.3527 & 2  \\
        2 & 1.7797 & -0.3403 & 2  \\
        3 & 1.9105 & -0.2095 & 2  \\
        4 & 1.9232 & -0.1968 & 2  \\
        \hline
    \end{tabular}
    \caption{Exciton spectrum from the tight-binding model for MoS$_2$ computed with $N_k=40^2$, $N_v=N_c=2$. This model has a direct gap at $\bold{K}$ of 2.12 eV, used to compute the shown exciton binding energies.}
\end{table}

After checking convergence with the number of $\bold{k}$ points and the number of bands, we obtain the spectrum shown in table (\ref{tab:mos2}). In this case, the point group of the material is again $D_{3h}$ 
and the irreducible representations realized by the wavefunctions at $\bold{Q}=0$ are compatible with the character table of the group \cite{bir1974symmetry}.
As before, to ensure that the excitons were computed correctly we can plot the total densities to ensure that they have the expected symmetries.

\begin{figure}[h]
    \centering
    \includegraphics[width=1\columnwidth]{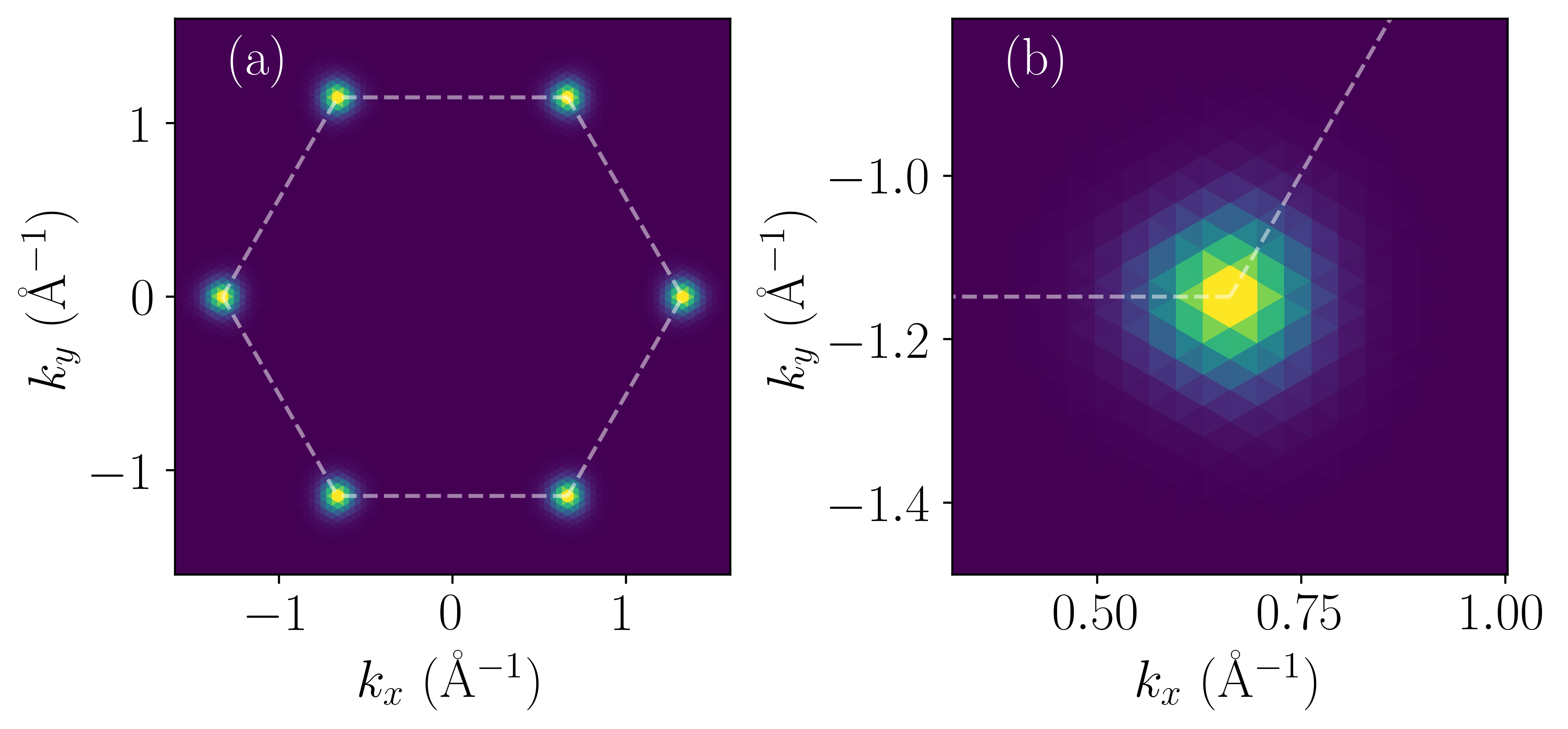}
    \caption{(a) Probability density of the ground state exciton in MoS$_2$ obtained over the full BZ with $N_k = 60^2$ for $\bold{Q}=0$. (b) Ground state exciton computed in a contour of the $\bold{K}$ valley with $N_k=30^2$ with a reduction factor of 2. Both calculations were done with $N_c=N_v=2$.}
    \label{fig:mos2_kwf}
\end{figure}

In Fig. (\ref{fig:mos2_kwf}a) we show the reciprocal probability density of the first energy level. As opposed to hBN, we observe that the states are strongly localized at the valleys. 
Resolving the degeneracy by labeling each exciton with the $C_3$ eigenvalues would result in each exciton localized in a different valley \cite{FengchengWu2015}. This shows that at least the low energy spectrum of $\text{MoS}_2$ can be studied at one valley instead of the whole BZ \cite{bieniek2020}.  This allows to get a more precise description of the exciton since one can use a more refined mesh. The wavefunction for the exciton obtained at one valley can be seen in Fig. (\ref{fig:mos2_kwf}b), using the code feature to reduce the BZ mesh by some integer factor. For higher excited states this does not hold since the states become more extended across the BZ, reaching both valleys.

\begin{figure}[h]
    \centering
    \includegraphics[width=0.9\columnwidth]{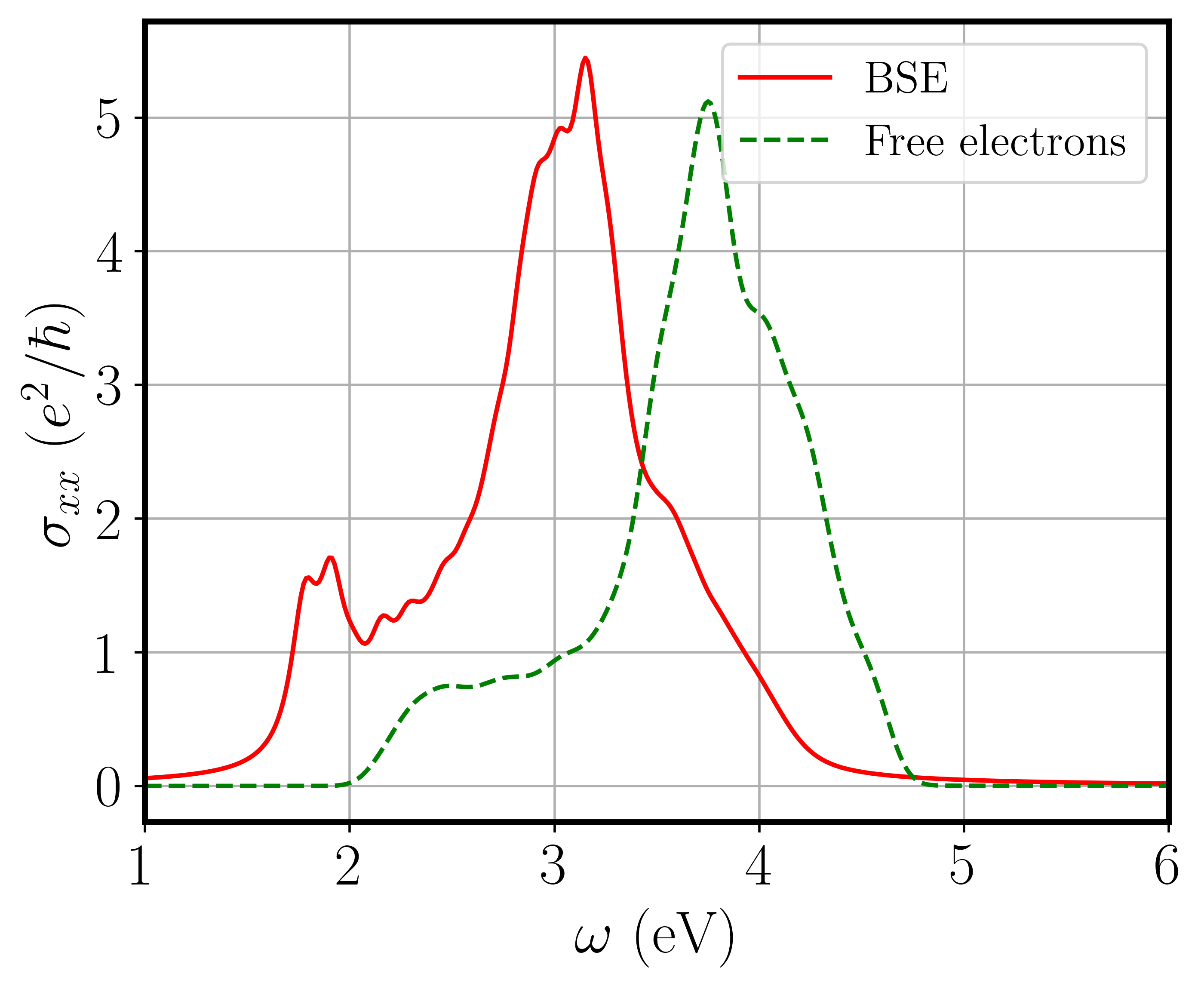}
    \caption{Optical conductivity of MoS$_2$ with and without excitons. The BSE calculation was done with $N_k=34^2$, $N_v=2$ and $N_c=6$. The first two peaks correspond to the A B excitons at the valleys, while the rest of the conductivity can be regarded as a shift of the non-interacting one.}
    \label{fig:mos2_absorption}
\end{figure}

Since the excitons are very localized in reciprocal space, they should be delocalized in real-space, meaning that the radius of the exciton should be large (e.g. compared to that of hBN). To complete the characterization of the excitons, we 
calculate the optical conductivity as shown in Fig. (\ref{fig:mos2_absorption})
While the exciton energies converge quickly with $N_k$, it is usually necessary to include more $\bold{k}$ points in the calculation of the optical conductivity in order to smooth unphysical oscillations derived from the discrete mesh. As it can be seen, the shape of the spectrum matches previous tight-binding studies \cite{Ridolfi2018,FengchengWu2015} and agrees well with ab-initio results \cite{qiu2013}. At low energies, the optical conductivity of MoS$_2$ presents the characteristic A and B exciton peaks, that are understood considering the main spin-allowed electron-hole excitations at the $\bold{K}$ and $\bold{K}'$ points. The split of $\sim$$100$ meV between such peaks reflects the effect of SOC in TMD materials \cite{liyilei2014}. At higher energies, the main feature of the spectra is a pronounced peak similar to the non-interacting case but red-shifted in energy. The excitons giving rise to such peak are often called ``C" excitons and were fully characterized in Ref. \cite{Ridolfi2018}, already showing the potential of tight-binding methods for studying new exciton physics.


\section{Conclusions}

We have developed a software package that allows to solve the Bethe-Salpeter equation constructed from either tight-binding models or DFT calculations based on localized orbitals. By considering orbitals as point-like, the computation of the interactions becomes  drastically simplified. Together with an effective screening, this results in a fast determination of the BSE matrix. More specifically, our real-space implementation of the interaction matrix elements is shown to be faster and more precise than its reciprocal-space counterpart, which is the formulation more commonly used.

As in GW-BSE approximations, the starting band structure plays a crucial role in determining the resulting exciton spectrum. Therefore it is key to select the best possible functional (typically hybrids) or the most accurate tight-binding models that capture the most prominent features of the band structure. Then, by choosing appropriately the screening parameters, it is possible to reproduce the results of GW-BSE or similar first-principles codes at a fraction of the computational cost. 

The Xatu code currently provides all the tools needed to extract and characterize the exciton spectrum, either using the binary or via its API. Nevertheless, the package is still under development, as new functionalities and optimizations are added. Our future plans include giving support for distributed parallelism to enable bigger system sizes and calculation of different excitation types such as trions or biexcitons. The code is currently aimed at the description of 2D materials, but it can support 0D and 3D systems. Since the Keldysh potential is only adequate for 2D systems, we will implement additional potentials suitable for different dimensionalities.
We also plan to add the possibility of performing exact calculations of the interaction matrix elements when using Gaussian-based DFT codes to compute the band structure. Currently we provide an interface with the CRYSTAL code \cite{crystal17}, and ideally more interfaces to community codes will be added over time, such as SIESTA \cite{siesta2020} or Wannier90 \cite{Pizzi_2020}. The project has been released under an open-source license and as such community contributions are welcome and encouraged. 

Note added: Upon completion of this work we became aware of a very recent submission which also addresses the problem of determining the exciton spectrum from Wannier-based tight-binding models \cite{DIAS2023}. Nevertheless, we believe that the thorough characterization of excitons we provide here can be advantageous and complementary to other different tools.

\section{Acknowledgments}
The authors acknowledge financial support from Spanish MICINN (Grant Nos. PID2019-109539GB-C43 \& TED2021-131323B-I00), María de Maeztu Program for Units of Excellence in R\&D (GrantNo.CEX2018-000805-M), Comunidad Autónoma de Madrid through the Nanomag COST-CM Program (GrantNo.S2018/NMT-4321), Generalitat Valenciana through Programa Prometeo (2021/017), Centro de Computación Científica of the Universidad Autónoma de Madrid, and Red Española de Supercomputación.

\begin{appendix}
\section{Symmetry}
\label{appendix:symmetry}

Throughout the document, we have mentioned and used several times the fact that each individual exciton state does not necessarily have the symmetries of the lattice, but it is the sum of the square amplitude of them who does. In this appendix we will give a proof of this statement: Given $|\Psi|^2=\sum_n|\psi_n|^2$, where $\psi_n$ denotes the wavefunction of the exciton states on some degenerate subspace of $PHP$, and given some symmetry operation $C$ such that $[H, C] = 0$, then
\begin{equation}
\label{symmetry_proof}
   C|\Psi|^2 = |\Psi|^2.
\end{equation}
First we have the consider the action of the symmetry operator $C$ on an exciton state. Given that the eigenstates of a degenerate subspace of $H$ are not in general eigenstates of $C$, the most general action is to mix the degenerate states, i.e.:
\begin{equation}
    C\psi_n=\sum_i\alpha_{in}\psi_i
\end{equation}
The coefficients $\alpha_{in}$ are the matrix elements of $C$. To prove (\ref{symmetry_proof}), we need to know the action of $C$ on the squared amplitude, $C|\psi_n|^2$. So first we want to prove the following property:
\begin{equation}
    C|\psi_n|^2=|C\psi_n|^2
\end{equation}
This can be proven using the action of the symmetry operation on the coordinate of the wavefunction, i.e. $C\psi_n(x) = \psi_n(C^{-1}x)$:
\begin{align}
    \nonumber C|\psi_n|^2(x)&=|\psi_n|^2(C^{-1}x)=\psi_n(C^{-1}x)\psi_n^*(C^{-1}x)\\
    &=C\psi_n(x)C\psi_n^*(x) = |C\psi_n|^2(x)
\end{align}
where we have also used that $C\psi_n^*(x) = (C\psi_n)^*(x)$. This last identity can be proved conjugating the action of the symmetry on the coordinates:
\begin{equation}
    (C\psi_n)^*(x) = \psi_n^*(C^{-1}x) = C\psi_n^*(x)
\end{equation}
This enables us to compute $C|\psi_n|^2$ in terms of a expansion on the states of the degenerate subspace:
\begin{equation}
    C|\psi_n|^2=|C\psi_n|^2=\left|\sum_i\alpha_{in}\psi_i\right|^2=\sum_{i,j}\alpha_{in}\alpha^*_{jn}\psi_i\psi^*_j
\end{equation}
Finally, with this expression we can prove the symmetry invariance of $|\Psi|^2 = \sum_n|\psi_n|^2$. To do so, we act with the symmetry operation $C$ on $|\Psi|^2$:
\begin{align}
    \nonumber C|\Psi^2| &= \sum_n C|\psi_n|^2 = \sum_n \left[\sum_{i,j}\alpha_{in}\alpha^*_{jn}\psi_i\psi^*_j\right] \\
    & = \sum_{ij}\left[\sum_n\alpha_{in}\alpha^*_{jn}\right]\psi_i\psi^*_j = \sum_{i}|\psi_i|^2 = |\Psi|^2
\end{align}
where we have used that $C$ is unitary, i.e. $\sum_n\alpha_{in}\alpha^*_{jn} = \delta_{ij}$. This proves that the sum of the squared amplitude of each degenerate state is invariant under the symmetry operations.
\\

On a different note, back in the examples we used the character table of the point group of the solid to justify the observed state degeneracies, and in the previous proof we also considered some general symmetry $C$ such that $[H, C] = 0$. For the abstract, unrepresented Hamiltonian, given any operation $C$ of the point group of the solid, it is true that $[H, C]= 0$. However, we are not working with the total Hamiltonian, but with a sector of it. So one must actually look for symmetry operations that commute with $PHP$:
\begin{equation}
    [PHP, C] = 0
\end{equation}
Since the sectors of electron-hole pairs of different momentum are disconnected, we can define $\Tilde{H}(\bold{Q}) = P_{\bold{Q}}HP_{\bold{\bold{Q}}}$, where $P_{\bold{Q}}$ is the projector over electron-hole pairs of $\bold{Q}$ total momentum. This Hamiltonian is analogous to the Bloch Hamiltonian $H(\bold{k})$, and it can be shown that it transforms in the same way:
\begin{equation}
    C^{-1}\Tilde{H}(\bold{Q})C = \Tilde{H}(C^{-1}\bold{Q}) 
\end{equation}
meaning that for $\bold{Q}=0$ the symmetry group is the crystallographic point group, but for $\bold{Q}\neq 0$ the Hamiltonian is invariant only under symmetry operations of the little group of $\bold{Q}$, whose irreducible representations thus dictate the (unitary) transformation properties of the $\bold{Q}-$excitonic wavefunctions.

Proof that $H(\bold{Q})$ transforms as the Bloch Hamiltonian $H(\bold{k})$ under symmetry operations:
\begin{align}
    \nonumber C^{-1}H(\bold{Q})C &= C^{-1}P_{\bold{Q}}CC^{-1}HCC^{-1}P_{\bold{Q}}C \\
    &= C^{-1}P_{\bold{Q}}CHC^{-1}P_{\bold{Q}}C 
\end{align}
where we have used that $[H,C]=0$. So we only need to see how the projectors transform under the symmetry operation to determine how $H(\bold{Q})$ transforms.
\begin{align} \label{PQtransformation}
    C^{-1}P_{\bold{Q}}C = \sum_{\bold{k},v,c}C^{-1}c^{\dagger}_{c,\bold{k+Q}}c_{v\bold{k}}\ket{GS}\bra{GS}c^{\dagger}_{v\bold{k}}c_{c,\bold{k+Q}}C
\end{align}
Inserting identities, we can transform each creation/annihilation operator according to $C^{-1}c^{\dagger}_{n\bold{k}}C = c^{\dagger}_{n,C^{-1}\bold{k}}$, up to an arbitrary phase that is cancelled in \eqref{PQtransformation}. From \eqref{FermiSea} it follows that the Fermi sea is invariant under point group operations, i.e. $C\ket{GS} = \ket{GS}$ (again, up to an arbitrary phase that is cancelled), we arrive to the following expression:
\begin{align}
    \nonumber &C^{-1}P_{\bold{Q}}C \\ 
    &= \sum_{\bold{k},v,c}c^{\dagger}_{c,C^{-1}\bold{k}+C^{-1}\bold{Q}}c_{v,C^{-1}\bold{k}}\ket{GS}\bra{GS}c^{\dagger}_{v,C^{-1}\bold{k}}c_{c,C^{-1}\bold{k}+C^{-1}\bold{Q}}
\end{align}
From the C-invariance of the BZ in $\bold{k}-$space, we arrive at the final expression for the transformed projector:
\begin{align}
    C^{-1}P_{\bold{Q}}C = \sum_{\bold{k},v,c}\ket{v,c,\bold{k},C^{-1}\bold{Q}}\bra{v,c,\bold{k},C^{-1}\bold{Q}} = P_{C^{-1}\bold{Q}}
\end{align}
Therefore, the projected exciton Hamiltonian $H(\bold{Q})$ also transforms in the same way:
\begin{align}
    C^{-1}H(\bold{Q})C = H(C^{-1}\bold{Q})
\end{align}
Likewise, the application of time-reversal yields $T^{-1}H(\bold{Q})T=H(-\bold{Q})$ whenever it is a symmetry of the system.

\section{Usage}
\label{section:usage}

The installation instructions can be found at the repository \url{https://github.com/alejandrojuria/xatu}, so it will not be discussed here.
The code has been developed with a hybrid approach in mind: one can resort to configuration files to run the program, in analogy with DFT codes, or instead program both the non-interacting system and run the exciton simulation using the provided API. First we are going to discuss its usage with configuration files. The basic usage as a CLI program is described by:
\begin{lstlisting}[backgroundcolor=\color{lightgray}, basicstyle={\small\ttfamily}]
xatu [OPTIONS] systemfile [excitonfile]
\end{lstlisting}
The executable always expects one file describing the system where we want to compute the excitons, and then another file specifying the parameters of the simulation. Their content is addressed in the next sections. The executable can also take optional flags, generally to tune the output of the simulation. By default, running the program without additional flags prints the exciton energies, without writing the results to any file. 
\begin{lstlisting}[backgroundcolor=\color{lightgray}, basicstyle={\small\ttfamily}]
-h (--help)
\end{lstlisting}
Used to print a help message with the usage of the executable and a list of all possible flags that may be passed. The simulation is not performed (even in presence of configuration files).
\begin{lstlisting}[backgroundcolor=\color{lightgray}, basicstyle={\small\ttfamily}]
-s (--states) nstates
\end{lstlisting}
The number of states specified with this flag is also used for any of the output flags. By default, the number of states is 8 (i.e. if the flag is not present).
\begin{lstlisting}[backgroundcolor=\color{lightgray}, basicstyle={\small\ttfamily}]
-p (--precision) decimals
\end{lstlisting}
One can specify the number of decimals used when printing the exciton energies. This is relevant to detect state degeneracy without inspecting manually the states. Defaults to 6 decimals if not present.
\begin{lstlisting}[backgroundcolor=\color{lightgray}, basicstyle={\small\ttfamily}]
-d (--dft) [ncells]
\end{lstlisting}
This flag is used to indicate that the \texttt{systemfile} provided corresponds to a CRYSTAL output file, instead of following the standarized format. DFT calculations usually involve several unit cells to determine the Bloch Hamiltonian, so the optional value \texttt{ncells} can be passed to specify how many we want to take into account. Otherwise all of them are read and used.
\begin{lstlisting}[backgroundcolor=\color{lightgray}, basicstyle={\small\ttfamily}]
-eck (--energy, --eigenstates, --kwf)
\end{lstlisting}
The optional flags \texttt{-e}, \texttt{-c}, \texttt{-k}, \texttt{-r} are used to specify which exciton output is written to file. \texttt{-e} writes the energies, \texttt{-c} writes the eigenvectors, \texttt{-k} writes the reciprocal density. Note that they can be combined instead of being written separately (e.g. \texttt{-ek} instead of \texttt{-e -k}).
\begin{lstlisting}[backgroundcolor=\color{lightgray}, basicstyle={\small\ttfamily}]
-r (--rswf) [holeIndex] [-r ncells]
\end{lstlisting}
Used to write the real-space probability densities to a file. One can give the index of the atom where the hole is located (defaults to first atom of the motif). It can be used a second time to specify the number of unit cells where we want to compute the amplitude (e.g. \texttt{-r 1 -r 10} fixes the hole at the second atom of the motif, and uses 10 unit cells along each axis). 
\begin{lstlisting}[backgroundcolor=\color{lightgray}, basicstyle={\small\ttfamily}]
-s (--spin)
\end{lstlisting}
Computes the total spin of the excitons, and writes it to a file. This assumes that the single-particle basis includes spin without performing any check, so incorrect usage could result in wrong results or runtime errors.
\begin{lstlisting}[backgroundcolor=\color{lightgray}, basicstyle={\small\ttfamily}]
-a (--absorption)
\end{lstlisting}
Computes the optical conductivity (which reflects the absorption of light up to a constant factor) as a function of frequency using the exciton spectrum, and saves the result to a file. A file named "kubo\_w.in" with the adequate format (shown below) must be present in the working directory.
\begin{lstlisting}[backgroundcolor=\color{lightgray}, basicstyle={\small\ttfamily}]
-m (--method) diag | davidson | sparse
\end{lstlisting}
Choose method to obtain the eigenstates of the BSE. By default, full diagonalization is used. If the Davidson or sparse (Lanczos) method is selected, then it is used to compute the number of states specified before.
\begin{lstlisting}[backgroundcolor=\color{lightgray}, basicstyle={\small\ttfamily}]
-b (--bands) kpointsfile
\end{lstlisting}
To check that the system file was written correctly, one can use this option to diagonalize the Bloch Hamiltonian on the $\bold{k}$ points specified on a file, and write the energy bands to a file. No exciton calculation is performed.

\subsection{Structure of a system file}

The system configuration files contain all the information needed to characterize completely the material of study: it provides the lattice vectors and motif positions, which is required for the real-space evaluation of the excitons. Then, we have the number of orbitals of each unique chemical species, which is needed to compute the matrix elements correctly, and the filling, which determines which bands participate in the formation of the exciton. Finally, the file contains the matrices needed to build the Bloch Hamiltonian, this is, the Fock matrices $H(\bold{R})$ and their corresponding Bravais vectors $\bold{R}$. The Bloch Hamiltonian is then reconstructed as:
\begin{equation}
\label{bloch_hamiltonian}
    H(\bold{k})=\sum_{\bold{R}}H(\bold{R})e^{i\bold{k}\cdot\bold{R}}
\end{equation}
Note that even though one has to provide the orbitals of each species, the specific type of orbital is not needed since the interaction is computed using the point-like approximation. A system file is specified using labels for each block. Blocks always begin with the block delimiter \texttt{\#}, followed by a label. A block is then defined as all the content between two consecutive block delimiters. The expected content for each label will be discussed next. Any line containing \texttt{!} is regarded as a comment, and empty lines are skipped.

\begin{itemize}
    \setlength{\itemindent}{-0.5cm}
    \item[\#] \texttt{BravaisLattice}: Basis vectors of the Bravais lattice. The number of vectors present is also used to determine the dimensionality of the system. The expected format is one vector per line, \texttt{x y z}.

    \item[\#] \texttt{Motif}: List with the positions and chemical species of all atoms of the motif (unit cell). The chemical species are specified with an integer index, used later to retrieve the number of orbitals of that species. The expected format is one atom per line, \texttt{x y z index}.

    \item[\#] \texttt{Orbitals}: Number of orbitals of each chemical species present. The position of the number of orbitals for each species follows the indexing used in the motif block. This block expects one or more numbers of orbitals, the same as the number of different species present, \texttt{n1 [n2 ...]}.

    \item[\#] \texttt{Filling}: Total number of electrons in the unit cell. Required to identify the Fermi level, which is the reference point in the construction of the excitons. Must be an integer number.

    \item[\#] \texttt{BravaisVectors}: List of Bravais vectors $\bold{R}$ that participate in the construction of the Bloch Hamiltonian (\ref{bloch_hamiltonian}). Expected one per line, in format \texttt{x y z}.

    \item[\#] \texttt{FockMatrices}: Matrices $H(\bold{R})$ that construct the Bloch Hamiltonian $H(\bold{k})$. The matrices must be fully defined, i.e., they cannot be triangular, since the code does not use hermiticity to generate the Bloch Hamiltonian. The Fock matrices given must follow the ordering given in the block \texttt{BravaisVectors}. The matrices can be real or complex, and each one must be separated from the next using the delimiter \texttt{\&}. In case the matrices are complex, the real and imaginary parts must be separated by a space, and the complex part must carry the imaginary number symbol (e.g. $1.5$  $-2.1j$). Both $i$ and $j$ can be used.

    \item[\#] \texttt{[OverlapMatrices]}: In case that the orbitals used are not orthonormal, one can optionally provide the overlap matrices $S(\bold{R})$. The overlap in $\bold{k}$ space is given by:
    $$S(\bold{k}) = \sum_{\bold{R}}S(\bold{R})e^{i\bold{k}\cdot\bold{R}}$$
    This is necessary to be able to reproduce the bands, which come from solving the generalized eigenvalue problem $H(\bold{k})S(\bold{k})\Psi = ES(\bold{k})\Psi$.
    This will be specially necessary if the system was determined using DFT, since in tight-binding we usually assume orthonormality. This block follows the same rules as \texttt{FockMatrices}: each matrix $S(\bold{R})$ must be separated with the delimiter \texttt{\&}, and they must follow the order given in \texttt{BravaisVectors}.
\end{itemize}

Several examples of valid system files are provided in the code repository, under the folder \texttt{/models}.

\subsection{Structure of an exciton file}
The purpose of the system file was to specify completely the system where we want to compute the excitons. Then, the exciton file is used to describe the excitons themselves: number of points in the mesh or submesh, bands that participate and center-of-mass momentum for example, as well as some additional flags. The idea is to keep the functionality as orthogonal as possible between files. With one system file, we can test for the convergence of the excitons with the number of kpoints, or with the number of bands modifying the exciton file only. Finally, we have the runtime options of the program, which in general do not affect the energy and modify the output exclusively. The philosophy is to maximize the reproducibility and facilitate tracking of the experiments.

The exciton files are built following the rules of the system files. They are composed of blocks, starting with \texttt{\#}. Each block has a label, which determines the expected content of the block. Next we provide a list of the possible parameters used in the construction of an exciton file:

\begin{itemize}
    \setlength{\itemindent}{-0.5cm}
    \item[\#] \texttt{Label}: Used to specify the name of the files containing the output of the program. The files will be named \texttt{[Label].eigval}, \texttt{[Label].kwf}, etc.
    
    \item[\#] \texttt{Bands}: Number of bands above and below the Fermi level. The minimum value is 1, to describe one conduction band and one valence band (i.e. only one combination of bands).

    \item[\#] \texttt{[BandList]}: As an alternative to \texttt{Bands}, one can specify a list with the indices of the bands that compose the exciton. 0 is taken as the last valence band, meaning that 1 would be the first conduction band, -1 is the second valence band and so on. This option can be used to generate asymmetric combinations of bands. It overrides the \texttt{Bands} block.

    \item[\#] \texttt{Ncells}: Number of points in one direction of the Brillouin zone, or equivalently number of unit cells along one axis. The same number of points is taken along all directions.

    \item[\#] \texttt{[Submesh]}: Used to specify a submesh of the Brillouin zone. Takes a positive integer $m$, which divides the BZ along each axis by that factor. The resulting area is meshed with the number of points specified in the \texttt{Ncells} block. This option can become memory intensive (it scales as $\mathcal{O}(m^d)$, $d$ the dimension).

    \item[\#] \texttt{[ShiftMesh]}: In case that we are using a submesh, then probably we also want to shift the meshed area to center it at the gap, where the exciton peaks. Takes a vector with its components, \texttt{kx ky kz}.

    \item[\#] \texttt{Dielectric}: The Keldysh interaction requires setting the dielectric constants of substrate $\epsilon_s$, the medium $\epsilon_m$ and the screening length $r_0$, which involves the dielectric constant of the material. This block expects three values, \texttt{es em r0}.

    \item[\#] \texttt{[TotalMomentum]}: One can optionally specify the total or center-of-mass momentum $\bold{Q}$ of the exciton. By default, it is taken to be zero, unless this block is specified. It expects a vector in form \texttt{qx qy qz}.

    \item[\#] \texttt{[Reciprocal]}: If present, the interaction matrix elements are computed in reciprocal space instead of direct space, which is the default. It takes an integer argument to specify the number of reciprocal cells to sum over, \texttt{nG}.

    \item[\#] \texttt{[Exchange]}: Flag to turn on the exchange interaction. By default computations neglect the exchange, and use only the direct term. It has to be set to \texttt{true} or \texttt{false}.

    \item[\#] \texttt{[Scissor]}: Used to specify a scissor shift of the bands to correct the gap. This optional field takes a single value, \texttt{shift}
\end{itemize}

As it can be seen, a minimum exciton simulation only requires specifying the number of bands, the number of $\bold{k}$ points and the dielectric constants. The modification of any of the present parameters is expected to result in a variation of the exciton results (energies, wavefunctions, conductivity), which is why all this parameters have been delegated to the same file.

\subsection{Absorption file}
For the calculation of the Kubo conductivity, one needs to provide a separate input file named \texttt{kubo\_w.in} in the working folder. This file is used to specify all parameters relative to the conductivity calculation, namely the desired energy interval, the point sampling and the broadening to be used, as well as the output files. Its format is as follows:
\begin{verbatim}
#initial frequency (eV)
5
#frequency range (eV)
8
#number of frequency points (integer)
300
#broadening parameter (eV)
0.05
#type of broadening
lorentzian
#output kubo name files
kubo_hBN_sp.dat
kubo_hBN_ex.dat
\end{verbatim}
Do note that as opposed to the previous configuration files, the name of each section starting with $\#$ is not actually relevant for the parsing; the program always expects the same fields to be present in the file, and in the same order as presented here. For the broadening, three different options are allowed (\texttt{'lorentzian', 'exponential', 'gaussian'})

\subsection{As a library}

So far we have discussed a more streamlined usage of the package. In some cases, however, the user could benefit from accessing directly the results of an exciton calculation, instead of having to dump it to a file to postprocess it later. To enable this possibility, the package has been also designed as a library, meaning that one can import the classes and functions defined in the API, and use them to build some extra functionality. Some use cases could be scenarios with exciton interactions, such as exciton-exciton interactions or exciton-polaritons.

To do so, the package provides a header file which defines a namespace. Within the namespace we have access to all the exciton functionality, which is completely documented. For instructions on how to build the documentation, we refer to the project repository where the most up-to-date information will be present. Additionally, some usage examples can be found under the root directory in the folder \texttt{/main}. 

The outline for a general exciton simulation is the following: one first has to create a System object, which can be done with a system file. Alternatively, one can define a subclass that inherits from System, and use it to implement the desired behaviour (namely the Bloch Hamiltonian). Then this System is passed on to the Exciton class, which we configure with the desired parameters. The interacting Hamiltonian is initialized and solved, returning a Result object which contains the eigenvalues and eigenvectors. With this, now we can compute some observables, or instead use these states to perform some other calculations out of the scope of the code.

\subsection{Output}

To conclude the usage section, we will describe briefly the structure of the output files. Here we describe how each file is written so the user can write their own custom routines; we also provide some example Python scripts under the folder \texttt{/plot}.

\begin{itemize}
    \item Energy: The energy file has in the first line the total number of energies written in the file. The second line contains all the energies, separated by a tabulation, \texttt{e1 e2 ... en}. All energies are written, including degenerate levels. All the exciton energies are given with respect to the Fermi sea energy. To obtain the binding energy, one must substract the gap from the exciton energy. Units are $[E]=$ eV.

    \item States: The first line contains the dimension of the BSE matrix $n$, i.e. the number of different electron-hole pairs. The next $n$ lines specify the valence, conduction bands of each electron-hole pair and their $\bold{k}$ point, \texttt{kx ky kz v c}. Afterwards, each line specifies completely the coefficients of each exciton state. The format per line is: \texttt{Re(A1) Im(A1) Re(A2) Im(A2) ...}.

    \item Reciprocal probability density: For the reciprocal density, on each line we specify the coordinates of the k point and the associated probability, \texttt{kx ky kz P}. Each state is separated from the next by a delimiter \texttt{\#}. Units are $[k] = $\AA$^{-1}$.

    \item Real-space probability density: The first line has the coordinates of the hole, \texttt{hx hy hz}. The following ones each have the coordinates of one atomic position, and the probability of finding the electron: \texttt{x y z P}. Densities for different states are separated by \texttt{\#}. Units are $[x]=$\AA.

    \item Spin: On each line we write the index of the current exciton, and next the total spin projection, the hole and the electron spin, \texttt{n St Sh Se}. Spin units are $[S_z]=\hbar$.

    \item Absorption: Both conductivities with and without exciton effects are computed and written to two different files. Each row contains the following columns:
    $\omega,\ \sigma_{xx},\ \sigma_{xy},\ \sigma_{xz}$,\ $\sigma_{yx},\ \sigma_{yy},\ \sigma_{yz},\ \sigma_{zx},\ \sigma_{zy},\ \sigma_{zz}$. Units are $[\omega] = $ eV, $[\sigma_{ij}] = e^2/\hbar$.
\end{itemize}

\end{appendix}





\bibliographystyle{elsarticle-num}
\bibliography{biblio.bib}







\end{document}